\documentclass[preprint,showpacs,amsmath,amssymb,aps,prd,nofootinbib]{revtex4}
\usepackage{epsfig,graphicx,color,appendix}

\begin{document}
\begin{flushright}
KIAS-P13033
\end{flushright}

\title{\mbox{}\\[10pt]
A dynamical CP source for CKM, PMNS \\and Leptogenesis}

\author{Y. H. Ahn\footnote{Email: yhahn@kias.re.kr},
Seungwon Baek\footnote{Email: swbaek@kias.re.kr}
}

\affiliation{School of Physics, KIAS, Seoul 130-722, Korea}


\begin{abstract}
We propose a model for the spontaneous CP violation based on $SU(2)_{L}\times U(1)_{Y}\times A_{4}\times CP\times Z_{2}$ symmetry for quarks and leptons in a seesaw framework. We investigate a link between the CP phase in the Cabibbo-Kobayashi-Maskawa (CKM) matrix and CP phase in the Pontecorvo-Maki-Nakagawa-Sakata (PMNS) matrix by using the present data of quark sector.
In our model CP is spontaneously broken at high energies, after breaking of flavor symmetry, by a
complex vacuum expectation value of $A_{4}$-triplet and gauge singlet scalar field.
And, certain effective dimension-5 operators are considered in the Lagrangian as an equal footing, in which the quarks lead to the CKM matrix of the quark mixing. However, the lepton Lagrangian still keep renormalizability, which gives rise to a non-degenerate Dirac neutrino Yukawa matrix, a unique CP-phase, and the nonzero value of $\theta_{13}\simeq9^{\circ}$ as well as two large mixing angles $\theta_{12}, \theta_{23}$.
We show that the generated CP phase ``$\xi$" from the spontaneous CP violation could become a natural source of leptogenesis as well as CP violations in the CKM and PMNS.
Interestingly enough, we show that, for around $\xi\simeq110^{\circ}(140^{\circ})$, we obtain the measured CKM CP-phase  $\delta^{q}_{CP}\simeq70^{\circ}$ for normal (inverted) hierarchy. For the measured value of $\theta_{13}$ we favor the PMNS CP-phase around $30^\circ, 200^\circ$, and $|\theta_{23}-45^{\circ}|\rightarrow0$ for normal mass hierarchy and around
$60^\circ, 110^\circ, 230^\circ$, $|\theta_{23}-45^{\circ}|\rightarrow5^{\circ}$ for inverted one.
As a numerical study in the lepton sector, we show low-energy phenomenologies and leptogenesis for the normal and inverted case, respectively, and a interplay between them.
\end{abstract}

\maketitle %
\section{Introduction}

CP violation (CPV) plays a crucial role in our understanding of the observed baryon asymmetry of the Universe (BAU)~\cite{Farrar:1993hn}.
This is because the preponderance of matter over antimatter in the observed Universe cannot be generated from an equal amounts of matter and antimatter unless CP is broken as  shown by Sakharov (1967), who pointed out that in addition to CP violation baryon-number violation, C (charge-conjugation) violation, and a departure from thermal equilibrium are all necessary to successfully achieve a net baryon asymmetry in early Universe.
In the Standard Model (SM)
CP symmetry is violated due to a complex phase in the Cabibbo-Kobayashi-Maskawa (CKM) matrix \cite{CKM}. However, since the extent of CP violation in the SM is not enough for achieving the observed BAU,  we need new source(s) of CP violation for a successful BAU.
On the other hand, CP violations in the lepton sector are imperative if the BAU could be realized through leptogenesis.
So, any hint or observation of the leptonic CP violation  can strengthen our belief in leptogenesis~\cite{review, Khlopov}.

The violation of the CP symmetry is a crucial ingredient of any dynamical mechanism which intends to explain both low energy CP violation and the baryon asymmetry.
Renormalizable gauge theories are based on the spontaneous symmetry breaking mechanism, and it is natural to have the spontaneous CP violation (SCPV)~\cite{Lee:1973iz, Branco:1999fs}
as an integral part of that mechanism. Determining all
possible sources of CP violation is a fundamental challenge for high energy physics.
In economical viewpoint, it would be good if both leptonic- and quark-sector CPV phases
are originated from a single source, for example, the one in the complex vacuum
in the SCPV~\cite{Lee:1973iz, Branco:1999fs}.
There is a common problem in models with SCPV, however, which is that a strong QCD
$\bar{\vartheta}_{\rm eff}$ term will be generated. However, a SCPV can provide a solution to the strong CP problem, if the parameter $\bar{\vartheta}_{\rm eff}$ related to the strong CP problem is vanishing at tree level and calculable at higher orders~\cite{SCPV}.

We propose a model for the SCPV based on an $A_{4}$ flavor symmetry for quarks and leptons in a seesaw framework. The seesaw mechanism, besides explaining of smallness of the measured neutrino masses,  has another appealing feature: generating the observed baryon asymmetry in our Universe by means of leptogenesis~\cite{review}.
CP symmetry is spontaneously broken at high energies, after breaking of $A_{4}$ flavor symmetry, by a complex vacuum expectation value (VEV) of $A_{4}$-triplet and gauge singlet scalar filed $\chi$, which is introduced to give the correct flavor structure in the heavy neutrino sector.
The main goal of our work is twofold: First, we investigate CP violation in the quark and
lepton sectors and show how CP phases in both CKM and Pontecorvo-Maki-Nakagawa-Sakata (PMNS) matrices
can be obtained simultaneously through spontaneous symmetry breaking mechanism. Second, we show that
the phase generated through the SCPV can be a unique CP source of both CKM and PMNS matrices, and discuss
how to link between leptonic mixing and leptogenesis through the SCPV.

This work is an extension of that in~\cite{Ahn:2012cg} in such a way that
(i) the $A_{4}$ flavor symmetry is spontaneously broken, and thereby a CP breaking phase is generated spontaneously, and (ii) in the quark sector all possible effective dimension-5 operators which are invariant under $SU(2)_L\times U(1)_Y\times A_{4}\times CP \times Z_{2}$ symmetry are introduced to explain the CKM matrix,
while in the lepton sector the renomalizability constraint is kept.
Thus, our model can naturally explain both the CKM mixing parameters (the three angles, $\theta^{q}_{23},\theta^{q}_{13},\theta^{q}_{12}$, and the CKM CP phase $\delta^{q}_{CP}$) and the PMNS mixing angles ($\theta_{23},\theta_{13},\theta_{12}$).

This paper is organized as follows. In the next section, we show the particle content
and its representations under the $A_4$  flavor symmetry and an auxiliary $Z_{2}\times CP$ symmetry in our model, as well as construct a Higgs scalar and a Yukawa Lagrangian. In Sec.~III, we discuss how to realize the spontaneous breaking of CP symmetry, and then we outline the minimization of the scalar potential and the vacuum alignments.
In Sec.~IV, we consider the phenomenology of quarks and leptons at low-energy, and in Sec.~V we study numerical analysis for neutrino oscillations and provide the data points for the CKM and PMNS. In Sec. VI we show possible leptogenesis and its link with low energy observables.
We give our conclusions in Sec.~VII.

\section{The Model}
 \label{Model}
In the absence of flavor symmetries, particle masses and mixings
are generally undetermined in a gauge theory. In order to understand the present data for quarks and leptons, especially, the CKM mixing angles ($\theta^{q}_{23},\theta^{q}_{13},\theta^{q}_{12}$ with the CKM CP-phase $\delta^{q}_{CP}$) and the nonzero $\theta_{13}$~\cite{Theta13, Ahn:2012tv} and tri-bimaximal mixing (TBM) angles~\cite{TBM} ($\theta_{12}, \theta_{23}$) for the neutrino oscillation data and baryogenesis via leptogenesis, as well as to predict a CP violation of the lepton sector, we propose a simple discrete symmetry model for the SCPV based on an $A_{4}$ flavor symmetry for quarks and leptons.
Here we recall that $A_{4}$ is the symmetry group of the tetrahedron and a finite group of even permutation of four objects~\cite{Ma:2001dn}.
The group $A_{4}$ has two generators $S$ and $T$, satisfying the relation $S^{2}=T^{3}=(ST)^{3}={\bf 1}$.
In the three-dimensional unitary representation, $S$ and $T$ are given by
 \begin{eqnarray}
 S={\left(\begin{array}{ccc}
 1 &  0 &  0 \\
 0 &  -1 & 0 \\
 0 &  0 &  -1
 \end{array}\right)}~,\qquad T={\left(\begin{array}{ccc}
 0 &  1 &  0 \\
 0 &  0 &  1 \\
 1 &  0 &  0
 \end{array}\right)}~.
 \label{generator}
 \end{eqnarray}
The group $A_{4}$ has four irreducible representations, one triplet ${\bf 3}$, and three singlets ${\bf 1}, {\bf 1}', {\bf 1}''$
with the multiplication rules ${\bf 3}\otimes{\bf 3}={\bf 3}_{s}\oplus{\bf 3}_{a}\oplus{\bf 1}\oplus{\bf 1}'\oplus{\bf 1}''$,
${\bf 1}'\otimes{\bf 1}''={\bf 1}$, ${\bf 1}'\otimes{\bf 1}'={\bf 1}''$,
and ${\bf 1}''\otimes{\bf 1}''={\bf 1}'$.
Let us denote two $A_4$ triplets as $(a_{1}, a_{2}, a_{3})$ and $(b_{1}, b_{2}, b_{3})$,
then we have
 \begin{eqnarray}
  (a\otimes b)_{{\bf 3}_{\rm s}} &=& (a_{2}b_{3}+a_{3}b_{2}, a_{3}b_{1}+a_{1}b_{3}, a_{1}b_{2}+a_{2}b_{1})~,\nonumber\\
  (a\otimes b)_{{\bf 3}_{\rm a}} &=& (a_{2}b_{3}-a_{3}b_{2}, a_{3}b_{1}-a_{1}b_{3}, a_{1}b_{2}-a_{2}b_{1})~,\nonumber\\
  (a\otimes b)_{{\bf 1}} &=& a_{1}b_{1}+a_{2}b_{2}+a_{3}b_{3}~,\nonumber\\
  (a\otimes b)_{{\bf 1}'} &=& a_{1}b_{1}+\omega a_{2}b_{2}+\omega^{2}a_{3}b_{3}~,\nonumber\\
  (a\otimes b)_{{\bf 1}''} &=& a_{1}b_{1}+\omega^{2} a_{2}b_{2}+\omega a_{3}b_{3}~,
 \end{eqnarray}
where $\omega=e^{i2\pi/3}$ is a complex cubic-root of unity.

To make the presentation of our model physically more transparent, we define the $T$-flavor quantum number $T_f$ as the eigenvalues of the operator $T$, for which $T^3=1$. In detail, we say that a field $f$ has $T$-flavor $T_f=0$, +1, or $-$1 when it is an eigenfield of the $T$ operator with eigenvalue $1$, $\omega$, $\omega^2$, respectively (in short, with eigenvalue $\omega^{T_f}$ for $T$-flavor $T_f$, considering the cyclical properties of the cubic root of unity $\omega$). The $T$-flavor is an additive quantum number modulo 3. We also define the $S$-flavor parity as the eigenvalues of the operator $S$, which are +1 and -1 since $S^2=1$, and we speak of $S$-flavor-even and $S$-flavor-odd fields.

We extend the SM by the inclusion of an $A_4$-triplet of
right-handed $SU(2)_{L}$-singlet Majorana neutrinos $N_{R}$, and the introduction of three types of scalar Higgs fields  besides the SM-like $SU(2)_{L}$-doublet Higgs bosons $\Phi$, which we take to be an $A_4$-triplet: a second $SU(2)_{L}$-doublet of Higgs bosons $\eta$, which is distinguished from $\Phi$ by being an $A_4$-singlet with no $T$-flavor (singlet representation), an $SU(2)_{L}$-singlet $A_4$-triplet scalar field $\chi$:
 \begin{eqnarray}
  \Phi_j =
\begin{pmatrix} \varphi^{+}_j \\ \varphi^{0}_j \end{pmatrix},
\qquad
\eta =
\begin{pmatrix}
  \eta^{+} \\
  \eta^{0}
\end{pmatrix},
\qquad
\chi_j,
\qquad
(j=1,2,3).
  \label{Higgs}
 \end{eqnarray}

We assign each flavor of both leptons and right-handed quarks to one of the three $A_4$ singlet representations: the electron ($u$, $d$-quark)-flavor to the ${\bf 1}$ ($T$-flavor 0), the muon ($c$, $s$-quark) flavor to the ${\bf 1}''$  ($T$-flavor -1), and the tau ($t$, $b$-quark) flavor to the ${\bf 1}'$  ($T$-flavor +1). And, we assign left-handed quarks $Q_{L}$ to the $A_{4}$ triplet representation. (Note in this respect that our $A_4$ flavor group is not a symmetry under exchange of any two lepton (quark) flavors, like $e$ and $\mu$, for example. Our $A_4$ flavor group is implemented as a global symmetry of the Lagrangian, later spontaneously broken, but some fields are not invariant under $A_4$ transformations, much in the same way as the implementation of $SU(2)_L\times U(1)_Y$ in the SM, where left-handed and right-handed fermions are assigned to different representations of the gauge group. Then we take the Higgs boson doublet $\eta$ to be invariant under $A_4$, that is to be a flavor-singlet ${\bf 1}$ with no $T$-flavor. The other Higgs doublet $\Phi$, the Higgs singlet $\chi$, and the singlet neutrinos $N_R$ are assumed to be triplets under $A_4$, and so can be used to introduce lepton-flavor violation in an $A_4$ symmetric Lagrangian.

The field content of our model and the field assignments to $SU(2)_L\times U(1)_Y\times A_{4}$ representations are summarized in Table~\ref{reps}.
\begin{table}[h]
\begin{widetext}
\begin{center}
\caption{\label{reps} Representations of the fields under $A_4$ and $SU(2)_L \times U(1)_Y \times Z_2$. }
\begin{ruledtabular}
\begin{tabular}{ccccccccccc}
Field &$L_{e},L_{\mu},L_{\tau}$&$Q_{L}$&$e_R,\mu_R,\tau_R$&$u_R,c_R,t_R$&$d_R,s_R,b_R$&$N_{R}$&$\chi$&$\eta$&$\Phi$\\
\hline
$A_4$&$\mathbf{1}$, $\mathbf{1^{\prime\prime}}$, $\mathbf{1^{\prime}}$&$\mathbf{3}$&$\mathbf{1}$, $\mathbf{1^{\prime\prime}}$, $\mathbf{1^{\prime}}$&$\mathbf{1}$, $\mathbf{1^{\prime\prime}}$,$\mathbf{1^{\prime}}$
 &$\mathbf{1}$, $\mathbf{1^{\prime\prime}}$,$\mathbf{1^{\prime}}$&$\mathbf{3}$&$\mathbf{3}$&$\mathbf{1}$&$\mathbf{3}$\\
$Z_{2}$&$-$&$+$&$+$&$+$
 &$+$&$-$&$+$&$-$&$+$\\
$SU(2)_L\times U(1)_Y$&$(2,-1)$&$(2,\frac{1}{3})$&$(1,-2)$&$(1,\frac{4}{3})$
 &$(1,-\frac{2}{3})$&$(1,0)$&$(1,0)$&$(2,1)$&$(2,1)$\\
\end{tabular}
\end{ruledtabular}
\end{center}
\end{widetext}
\end{table}
In addition to $A_{4}$ flavor symmetry, we impose an additional symmetry $Z_{2}$, where $L_{e,\mu,\tau}$, $N_{R}$, and $\eta$ carries $Z_{2}$-odd quantum number, while all other fields have a $Z_{2}$-even one.
So this non-flavor symmetry forbids some irrelevant $SU(2)_{L}\times U(1)_{Y}\times A_{4}$ invariant Yukawa terms from the Lagrangian (see the quark Lagrangian).

We impose $A_{4}$ flavor symmetry for leptons, quarks, and scalars, and force CP to be invariant at the Lagrangian level, which implies that all the parameters appearing in the Lagrangian are real. The extended Higgs sector can spontaneously break CP through a phase in the VEV of the singlet scalar field~\cite{Ahn:2013mva}. The CP invariance in the Lagrangian can be clarified by the nontrivial transformation~\cite{Holthausen:2012dk}
\begin{align}
\psi\rightarrow U\psi^{\ast}=\psi~,
\end{align}
where the $A_{4}$ triplet fields $\psi=N_{R}, \chi, \Phi$  and
\begin{align}
U={\left(\begin{array}{ccc}
 1 &  0 &  0 \\
 0 & 0 & 1 \\
 0 & 1 & 0
 \end{array}\right)}.
\end{align}

In our Lagrangian, we assume that there is a cutoff scale $\Lambda$, above which there exists unknown physics.

\subsection{Higgs sector}
The full quartic $SU(2)_{L}\times U(1)_{Y}\times A_{4}\times Z_{2}\times CP$ invariant Higgs potential in $\Phi,\eta$ and $\chi$ is displayed, in general, as
 \begin{eqnarray}
 V=V(\Phi)+V(\eta)+V(\chi)+V(\Phi\eta)+V(\Phi\chi)+V(\eta\chi)~,
\label{poten}
\end{eqnarray}
where
 \begin{eqnarray}
V(\Phi) &=& \mu^{2}_{\Phi}(\Phi^{\dag}\Phi)_{\mathbf{1}}+\lambda^{\Phi}_{1}(\Phi^{\dag}\Phi)_{\mathbf{1}}(\Phi^{\dag}\Phi)_{\mathbf{1}}+\lambda^{\Phi}_{2}(\Phi^{\dag}\Phi)_{\mathbf{1^\prime}}(\Phi^{\dag}\Phi)_{\mathbf{1^{\prime\prime}}}+\lambda^{\Phi}_{3}(\Phi^{\dag}\Phi)_{\mathbf{3}_{s}}(\Phi^{\dag}\Phi)_{\mathbf{3}_{s}}\nonumber\\
  &+&\lambda^{\Phi}_{4}(\Phi^{\dag}\Phi)_{\mathbf{3}_{a}}(\Phi^{\dag}\Phi)_{\mathbf{3}_{a}}+\lambda^{\Phi}_{5}\left\{(\Phi^{\dag}\Phi)_{\mathbf{3}_{s}}(\Phi^{\dag}\Phi)_{\mathbf{3}_{a}}+\text{h.c.}\right\}~,\\
V(\eta) &=& \mu^{2}_{\eta}(\eta^{\dag}\eta)+\lambda^{\eta}(\eta^{\dag}\eta)^{2}~,\\
V(\eta\Phi) &=& \lambda^{\eta\Phi}_{1}(\Phi^{\dag}\Phi)_{\mathbf{1}}(\eta^{\dag}\eta)+\lambda^{\eta\Phi}_{2}\left\{[(\Phi^{\dag}\eta)(\Phi^{\dag}\eta)]_{\mathbf{1}}+\text{h.c.}\right\}~,
\end{eqnarray}
 \begin{eqnarray}
V(\chi) &=& \mu^{2}_{\chi}\left\{(\chi\chi)_{\mathbf{1}}+\text{h.c.}\right\}+m^{2}_{\chi}(\chi\chi^{\ast})_{\mathbf{1}}+\left\{\lambda^{\chi}_{1}(\chi\chi)_{\mathbf{1}}(\chi\chi)_{\mathbf{1}}+\lambda^{\chi}_{11}(\chi^{\ast}\chi)_{\mathbf{1}}(\chi\chi)_{\mathbf{1}}+\text{h.c.}\right\}\nonumber\\
&+&\lambda^{\chi}_{12}(\chi^{\ast}\chi^{\ast})_{\mathbf{1}}(\chi\chi)_{\mathbf{1}}+
            \Big\{\lambda^{\chi}_{2}(\chi\chi)_{\mathbf{1}^\prime}(\chi\chi)_{\mathbf{1}^{\prime\prime}}+\lambda^{\chi}_{21} (\chi^{\ast}\chi)_{\mathbf{1}^\prime}(\chi\chi)_{\mathbf{1}^{\prime\prime}}+\lambda^{\chi}_{22} (\chi^{\ast}\chi)_{\mathbf{1}^{\prime\prime}}(\chi\chi)_{\mathbf{1}^{\prime}}\nonumber\\
&+&\lambda^{\chi}_{23}
            (\chi^{\ast}\chi^{\ast})_{\mathbf{1}^\prime}(\chi\chi)_{\mathbf{1}^{\prime\prime}}
  +\lambda^{\chi}_{3}(\chi\chi)_{\mathbf{3}_{s}}(\chi\chi)_{\mathbf{3}_{s}}+\lambda^{\chi}_{31}(\chi^{\ast}\chi)_{\mathbf{3}_{s}}(\chi\chi)_{\mathbf{3}_{s}}\nonumber\\
  &+&\lambda^{\chi}_{4}(\chi^\ast\chi)_{\mathbf{3}_{a}}(\chi\chi)_{\mathbf{3}_{s}}+\lambda^{\chi}_{41}(\chi\chi^\ast)_{\mathbf{3}_{a}}(\chi\chi)_{\mathbf{3}_{s}}+
  \xi^{\chi}_{1}\chi(\chi\chi)_{\mathbf{3}_{s}}+\xi^{\chi}_{11}\chi(\chi\chi^{\ast})_{\mathbf{3}_{s}}\nonumber\\
  &+&\xi^{\chi}_{12}\chi(\chi^{\ast}\chi^{\ast})_{\mathbf{3}_{s}}+\xi^{\chi}_{13}\chi(\chi\chi^{\ast})_{\mathbf{3}_{a}}+\text{h.c.}\Big\}+\lambda^{\chi}_{32}
            (\chi^{\ast}\chi^{\ast})_{\mathbf{3}_{s}}(\chi\chi)_{\mathbf{3}_{s}}~,
\end{eqnarray}
in which $V(\Phi\chi)$ and $V(\eta\chi)$ are given in Eqs.~(\ref{potential2}) and (\ref{potential3}).
Here $\mu_{\chi}$, $m_{\chi}$, $\xi^{\chi}_{1}$, $\xi^{\chi}_{11,12,13}$, $\mu_{\Phi}$ and $\mu_{\eta}$ have mass dimension-1, while $\lambda^{\chi}_{1,...,4}$, $\lambda^{\chi}_{11,12,21,22,23,31,32,,41}$, $\lambda^{\Phi}_{1,...,5}$, $\lambda^{\eta}$, and $\lambda^{\eta\Phi}_{1,2}$ are  dimensionless. In $V(\eta\Phi)$ the usual mixing term $\Phi^{\dag}\eta$ is forbidden by the $A_{4}$ symmetry.
In the presence of two $A_{4}$-triplet Higgs scalars $\chi$ and $\Phi$, Higgs potential terms involving both $\chi$ and $\Phi$, which would be written as $V(\Phi\chi)$ in Eq.~(\ref{poten}), would be problematic for vacuum stability and one could not get a desirable solution.
On the problematic Higgs potential $V(\Phi\chi)$ in Eq.~(\ref{potential3}), unnatural fine-tuning conditions are necessary for vacuum stability~\footnote{Such stability problems can be naturally solved, for instance, in the presence of a discrete symmetry~\cite{Ahn:2013mva}
or extra dimensions or in supersymmetric dynamical completions~\cite{A4, vacuum}. In these cases, $V(\Phi\chi)$ is not allowed or highly suppressed.}. In the limit where the seesaw scale $\chi$ field decouples from the electroweak Higgs fields $\Phi$ and  $\eta$, the decoupling of $\chi$ is performed by
 \begin{eqnarray}
  \lambda^{\Phi\chi}_{1,..,5, 11,21,31,41,51}\rightarrow0~,\qquad\xi^{\Phi\chi}_{1,2}\rightarrow0~,\qquad\lambda^{\eta\chi}_{1,2}\rightarrow0~.
  \label{Decouple}
 \end{eqnarray}
We wish these couplings to be exactly zero or sufficiently small,  where ``sufficiently small" means that those terms could not deform a demanded VEV alignment (see later).
Note that the potential $V(\eta\chi)$ does not affect demanded VEV alignments.

\subsection{Lepton sector}
The Yukawa interactions ($d\leq5$) in the neutrino and charged lepton sectors invariant under $SU(2)_L\times U(1)_Y\times A_4\times Z_{2}\times CP$ (including a Majorana mass term for the right-handed neutrinos) can be written as
 \begin{eqnarray}
 -{\cal L}^{\nu\ell}_{\rm Yuk} &=& y^{\nu}_{1}\bar{L}_{e}(\tilde{\Phi}N_{R})_{{\bf 1}}+y^{\nu}_{2}\bar{L}_{\mu}(\tilde{\Phi}N_{R})_{{\bf 1}''}+y^{\nu}_{3}\bar{L}_{\tau}(\tilde{\Phi}N_{R})_{{\bf 1}'}\nonumber\\
 &+&\frac{M}{2}(\overline{N^{c}_{R}}N_{R})_{{\bf 1}}+\frac{1}{2}y_R^\nu(\overline{N^{c}_{R}}N_{R})_{{\bf 3}_{s}} \chi\nonumber\\
 &+& y_{e}\bar{L}_{e}\eta~e_{R}+y_{\mu}\bar{L}_{\mu}\eta~\mu_{R}+y_{\tau}\bar{L}_{\tau}\eta~ \tau_{R}+\text{h.c.}~,
 \label{lagrangian}
 \end{eqnarray}
where $\tilde{\Phi}\equiv i\tau_{2}\Phi^{\ast}$ and $\tau_{2}$ is a Pauli matrix. Note here that there are no dimension-5 operators driven by $\chi$ field in the neutrino sector, and the above Lagrangian is renormalizable. The representation assignments and the requirement that the Lagrangian be renormalizable and $A_4$-symmetry forbid the presence of tree-level leptonic flavor-changing charged currents.
In this Lagrangian, each flavor of neutrinos and each flavor of charged leptons has its own independent Yukawa term, since they belong to different singlet representations ${\bf 1}$, ${\bf 1}''$, and ${\bf 1}'$ of $A_4$: the neutrino Yukawa terms involve the $A_{4}$-triplets $\Phi$ and $N_R$, which combine into the appropriate singlet representation; the charged-lepton Yukawa terms involve the $A_{4}$-singlet $\eta$ and the $A_4$-singlet right-handed charged-leptons $e_R$, $\mu_R$, and $\tau_R$. The right-handed neutrinos have an additional Yukawa term that involves the $A_4$-triplet SM-singlet Higgs $\chi$. The mass term $\frac{1}{2}M(\overline{N^{c}_{R}}N_{R})_{{\bf 1}}$ for the right-handed neutrinos is necessary to implement the seesaw mechanism by making the right-handed neutrino mass parameter $M$ large.

After electroweak and $A_4$ symmetry breaking, the neutral Higgs fields acquire vacuum expectation values and give masses to the charged-leptons and neutrinos: the Higgs doublet $\eta$ gives Dirac masses to the charge leptons, the Higgs doublet $\Phi$ gives Dirac masses to the three SM neutrinos, and the Higgs singlets $\chi$ gives Majorana masses to the right-handed neutrino $N_R$.
The charged lepton mass matrix is automatically diagonal due to the $A_{4}$-singlet nature of the charged lepton and Higgs field. The right-handed neutrino mass has the (large) Majorana mass contribution  $M$  and a contribution  induced by the electroweak-singlet $A_4$-triplet Higgs boson $\chi$ when the $A_4$-symmetry is spontaneously broken.

\subsection{Quark sector}
In the quark sector, the Yukawa interactions including dimension-5 operators $(d\leq5)$
driven by the $\chi$ field, invariant under $SU(2)_L \times U(1)_Y \times A_{4}\times Z_{2}\times CP$,
are given by
 \begin{eqnarray}
 {\cal L}^{q}_{\rm Yuk} = {\cal L}^{u}_{\rm Yuk} +{\cal L}^{d}_{\rm Yuk}~,
 \label{lagrangianCh}
 \end{eqnarray}
where
 \begin{eqnarray}
 -{\cal L}^{u}_{\rm Yuk} &=& y_{u}(\bar{Q}_{L}\tilde{\Phi})_{{\bf 1}}u_{R}
  +y_{c}(\bar{Q}_{L}\tilde{\Phi})_{{\bf 1}'}c_{R}
  +y_{t}(\bar{Q}_{L}\tilde{\Phi})_{{\bf 1}''}t_{R}\nonumber\\
 &+& \frac{y^{s}_{u}}{\Lambda}[(\bar{Q}_{L}\tilde{\Phi})_{{\bf 3}_{s}}\chi]u_{R}
  +\frac{y^{s}_{c}}{\Lambda}[(\bar{Q}_{L}\tilde{\Phi})_{{\bf 3}_{s}}\chi]_{{\bf 1}'}c_{R}
  +\frac{y^{s}_{t}}{\Lambda}[(\bar{Q}_{L}\tilde{\Phi})_{{\bf 3}_{s}}\chi]_{{\bf 1}''}t_{R}\nonumber\\
 &+& i\frac{y^{a}_{u}}{\Lambda}[(\bar{Q}_{L}\tilde{\Phi})_{{\bf 3}_{a}}\chi]u_{R}
  +i\frac{y^{a}_{c}}{\Lambda}[(\bar{Q}_{L}\tilde{\Phi})_{{\bf 3}_{a}}\chi]_{{\bf 1}'}c_{R}
  +i\frac{y^{a}_{t}}{\Lambda}[(\bar{Q}_{L}\tilde{\Phi})_{{\bf 3}_{a}}\chi]_{{\bf 1}''}t_{R}+ {\rm h.c.}~,  \label{lagrangianCh1}
 \end{eqnarray}
 \begin{eqnarray}
 -{\cal L}^{d}_{\rm Yuk} &=&
  y_{d}(\bar{Q}_{L}\Phi)_{{\bf 1}}d_{R}+y_{s}(\bar{Q}_{L}\Phi)_{{\bf 1}'}s_{R}
  +y_{b}(\bar{Q}_{L}\Phi)_{{\bf 1}''}b_{R}\nonumber\\
 &+&\frac{y^{s}_{d}}{\Lambda}[(\bar{Q}_{L}\Phi)_{{\bf 3}_{s}}\chi]d_{R}
  +\frac{y^{s}_{s}}{\Lambda}[(\bar{Q}_{L}\Phi)_{{\bf 3}_{s}}\chi]_{{\bf 1}'}s_{R}
  +\frac{y^{s}_{b}}{\Lambda}[(\bar{Q}_{L}\Phi)_{{\bf 3}_{s}}\chi]_{{\bf 1}''}b_{R}\nonumber\\
 &+&i\frac{y^{a}_{d}}{\Lambda}[(\bar{Q}_{L}\Phi)_{{\bf 3}_{a}}\chi]d_{R}
  +i\frac{y^{a}_{s}}{\Lambda}[(\bar{Q}_{L}\Phi)_{{\bf 3}_{a}}\chi]_{{\bf 1}'}s_{R}
  +i\frac{y^{a}_{b}}{\Lambda}[(\bar{Q}_{L}\Phi)_{{\bf 3}_{a}}\chi]_{{\bf 1}''}b_{R}+ {\rm h.c.}~.
 \label{lagrangianCh2}
 \end{eqnarray}
Note that in the above Lagrangian in order to keep CP invariance the imaginary ``$~i~$" is added in the terms associated with the antisymmetric product of two $A_4$ triplets in dimension-5 operators.
In the above Lagrangian, each flavor of up-type and down-type quarks has its own independent Yukawa term, since they belong to different singlet representations ${\bf 1}$, ${\bf 1}''$, and ${\bf 1}'$ of $A_4$: the terms involve the $A_{4}$-triplets $\Phi$ and $Q_{L}$, which combine into the appropriate singlet representation. The left-handed quark doublet $Q_{L}$
transforms as a triplet ${\bf 3}$, while the right-handed quarks (up-,down-type) $(u_{R}, d_{R})$,
$(c_{R}, s_{R})$, $(t_{R}, b_{R})$ transform as ${\bf 1}$, ${\bf 1}''$ and ${\bf 1}'$,
respectively.
We note that the $A_4$-triplet scalar field $\chi$ drives the dimension-5 operators in the quark sector shown in Eqs.~(\ref{lagrangianCh1}) and (\ref{lagrangianCh2}); the dimension-5 operator terms involve the $A_{4}$-triplet $\Phi$ and $Q_{L}$ fields, which combine into the right-handed quarks $u_R(d_R)$, $c_R(s_{R})$, and $t_R(b_{R})$. Thus, through spontaneous CP breaking this $\chi$ field plays a role to connect the lepton and quark sectors
to one another through the higher dimensional operators.

After electroweak and $A_4$ symmetry breaking, the neutral Higgs fields acquire vacuum expectation values and give masses to the up- and down-type quarks. In the renormalizable terms the Higgs doublet $\Phi$ gives Dirac masses to the up- and down-type quarks, and the quark mass matrices are automatically diagonal due to the $A_{4}$ structure of field contents; it provides the CKM matrix to be the identity, {\it i.e.} $V_{\rm CKM}=\textbf{I}$. Including higher dimensional operators driven by the Higgs singlets $\chi$ field give next-leading order masses to the up- and down-type quarks, and provide the correct CKM matrix (see later).

\section{Spontaneous CP violation}
 \label{SPCP}
The Higgs potential of our model is listed in Eq.~(\ref{poten01}). While CP symmetry is conserved at the Lagrangian level because all the parameters are assumed to be real, in our model  it can be spontaneously broken when both the $A_4$-triplets $\chi$ and $\Phi$ and the $A_4$-singlet $\eta$ acquire complex VEVs.
In addition, when a non-Abelian discrete symmetry like our $A_4$ is considered, it is crucial to check the stability of the vacuum.

Now let us discuss the how of realization of the spontaneous breaking of CP symmetry.
\subsection{Minimization of the neutral scalar potential}
The model contains four Higgs doublets and three Higgs singlets.
After electroweak- and $A_4$-symmetry breaking, we can find minimum configuration of the Higgs potential by taking the following:
 \begin{eqnarray}
  \langle\Phi_{j}\rangle &=&
  {\left(\begin{array}{c}
  0 \\
  \frac{e^{i\phi_j}}{\sqrt{2}}v_{\Phi_j}
 \end{array}\right)}~,\qquad \langle\eta\rangle =
  {\left(\begin{array}{c}
  0 \\
  \frac{e^{i\varphi}}{\sqrt{2}}v_{\eta}
 \end{array}\right)}~,\qquad\langle\chi_{j}\rangle=v_{\chi_j}e^{i\xi_{j}}~,
  \label{Higgs_vev}
 \end{eqnarray}
with $j=1,2,3$, where $v_{\Phi_{1,2,3}},v_{\chi_{1,2,3}}$, and $v_{\eta}$ are real and positive, and $\phi_{1,2,3},\xi_{1,2,3}$, and $\varphi$ are physically meaningful phases. The relative phases of $\Phi_{j}$, $\chi_{j}$, and $\eta$ are dynamically determined by minimizing the Higgs potential.

Since the $SU(2)_{L}$-singlet scalar field $\chi$ is much heavier than the other two gauge doublet scalar fields $\Phi$ and $\eta$, then the field $\chi$ is decoupled from the theory at an energy scale much higher than the electroweak scale.
In order for vacuum stability to be well described (see Eq.~(\ref{Decouple})), we assume more precisely
 \begin{eqnarray}
 \left|{\rm coupling~constants}(\Phi\chi)\right|\times\frac{|\rm seesaw~VEV|}{|\rm electroweak~VEV|}\ll1~.
\label{couple}
\end{eqnarray}
And, even the potential $V(\eta\chi)$ does not deform a desirable vacuum alignment, without loss of generality, here we have switched off the couplings in $V(\eta\chi)$.
Under the above assumptions, we get
 \begin{eqnarray}
 V= V(\chi)+V(\Phi)+V(\eta)+V(\eta\Phi)~.
\label{poten01}
\end{eqnarray}

First, the vacuum configuration for $\chi$ is obtained by vanishing the derivative of $V$ with respect to each component of the scalar fields $\chi_j$ and $\xi_j$. Then, we have three minimization equations
for VEVs and three equations for phases:
 \begin{eqnarray}
  \frac{\partial V}{\partial \chi_{j}}\Big|_{\chi_j=\langle\chi_j\rangle}=0~,\qquad\frac{\partial V}{\partial \xi_{j}}\Big|_{\xi_j}=0~,\qquad{\rm for}~j=1,2,3~.
 \end{eqnarray}
Concerning the above equations, by excluding the trivial solution where all VEVs vanish, we find
 \begin{eqnarray}
  \upsilon^{2}_{\chi}&=&-\frac{2m^{2}_{\chi}+4\mu^{2}_{\chi}\cos2\xi}
  {4(\tilde{\lambda}^{\chi}_{1}+2\tilde{\lambda}^{\chi}_{2}\cos2\xi+2(\lambda^{\chi}_{1}+\lambda^{\chi}_{2})\cos4\xi)}\neq0~,\quad
  \frac{\partial V}{\partial \chi_{2}}\Big|= \frac{\partial V}{\partial \chi_{3}}\Big|=0~,
  \label{vevChi}
 \end{eqnarray}
where $\tilde{\lambda}^{\chi}_{1}=\lambda^{\chi}_{12}+2\lambda^{\chi}_{23}$,
$\tilde{\lambda}^{\chi}_{2}=\lambda^{\chi}_{11}+\lambda^{\chi}_{21}+\lambda^{\chi}_{22}$ and $\langle\chi_{1}\rangle=v_{\chi}$. With the vacuum alignment of the $\chi$ field, Eq.~(\ref{vevChi}), the minimal condition for $\xi_{1}\equiv\xi$ is given as
 \begin{eqnarray}
  -\frac{1}{4}\frac{\partial V}{\partial \xi_{1}}\Big|_{\xi}&=&  v^{2}_{\chi}\left\{\mu^{2}_{\chi}+v^{2}_{\chi}\left(\tilde{\lambda}^{\chi}_{2}+4(\lambda^{\chi}_{1}+\lambda^{\chi}_{2})\cos2\xi\right)\right\}\sin2\xi=0~,
 \end{eqnarray}
and $\frac{\partial V}{\partial \xi_2}\Big|= \frac{\partial V}{\partial \xi_3}\Big|=0$ is automatically satisfied with respect to $\xi_{2}$, $\xi_{3}$.
So, we find a nontrivial VEV configuration for the $\chi$ field
 \begin{eqnarray}
  \langle\chi\rangle=v_{\chi}e^{i\xi}(1,0,0)~.
  \label{Alin}
 \end{eqnarray}
For the vacuum alignment given in Eq.~(\ref{Alin}), the scalar potential can be written as
 \begin{eqnarray}
  V_{\chi}&=&v^{4}_{\chi}\left\{\tilde{\lambda}^{\chi}_{1}+2\tilde{\lambda}^{\chi}_{2}\cos2\xi+2(\lambda^{\chi}_{1}+\lambda^{\chi}_{2})\cos4\xi\right\}
  +\frac{1}{2}v^{2}_{\chi}\left\{2m^{2}_{\chi}+4\mu^{2}_{\chi}\cos2\xi\right\}~.
 \end{eqnarray}
Depending on the values of $\xi$, the VEV configurations are given by:\\
{\bf (i)} for $\xi=0, \pm\pi$
 \begin{eqnarray}
  \upsilon^{2}_{\chi}&=&-\frac{2m^{2}_{\chi}+4\mu^{2}_{\chi}}
  {4(\tilde{\lambda}^{\chi}_{1}+2\tilde{\lambda}^{\chi}_{2}+2(\lambda^{\chi}_{1}+\lambda^{\chi}_{2}))}~,
  \label{VP1}
 \end{eqnarray}
{\bf (ii)} for $\xi=\pm\pi/2$
 \begin{eqnarray}
  \upsilon^{2}_{\chi}&=&\frac{-2m^{2}_{\chi}+4\mu^{2}_{\chi}}
  {4(\tilde{\lambda}^{\chi}_{1}-2\tilde{\lambda}^{\chi}_{2}+2(\lambda^{\chi}_{1}+\lambda^{\chi}_{2}))}~,
  \label{VP2}
 \end{eqnarray}
{\bf (iii)} for $\cos2\xi=-\frac{\mu^{2}_{\chi}+v^{2}_{\chi}\tilde{\lambda}^{\chi}_{2}}{4v^{2}_{\chi}(\lambda^{\chi}_{1}+\lambda^{\chi}_{2})}$
 \begin{eqnarray}
  \upsilon^{2}_{\chi}&=&\frac{2m^{2}_{\chi}(\lambda^{\chi}_{1}+\lambda^{\chi}_{2})-\tilde{\lambda}^{\chi}_{2}\mu^{2}_{\chi}}{\tilde{\lambda}^{\chi2}_{2}-4(\lambda^{\chi}_{1}+\lambda^{\chi}_{2})(\tilde{\lambda}^{\chi}_{1}-2(\lambda^{\chi}_{1}+\lambda^{\chi}_{2}))}~.
  \label{VP3}
 \end{eqnarray}
In the first case (i) the vacuum configurations do not violate CP, while the second (ii) and third case (iii) lead not only to the the spontaneous breaking of the CP symmetry but also to a nontrivial CP violating phase in the one loop diagrams relevant for leptogenesis.

Let us examine which case corresponds to the global minimum of the potential in a wide region of the parameter space.
Imposing the parameter conditions, $m^{2}_{\chi}<0$, $\tilde{\lambda}^{\chi}_{1,2}>0$ and $\lambda^{\chi}_{1,2}<0$, into Eqs.~(\ref{VP1}-\ref{VP3}),
the vacuum configurations of each case become
we obtain for the case (i)
 \begin{eqnarray}
  V_{0}=-\frac{(m^{2}_{\chi}+2\mu^{2}_{\chi})^2}{4\tilde{\lambda}^{\chi}_{1}+8(\tilde{\lambda}^{\chi}_{2}+\lambda^{\chi}_{1}+\lambda^{\chi}_{2})}~,\qquad \xi=0,\pm\pi~,
 \end{eqnarray}
for the case (ii)
 \begin{eqnarray}
  V_{0}=-\frac{(m^{2}_{\chi}-2\mu^{2}_{\chi})^2}{4\tilde{\lambda}^{\chi}_{1}-8(\tilde{\lambda}^{\chi}_{2}-\lambda^{\chi}_{1}-\lambda^{\chi}_{2})}~,\qquad \xi=\pm\frac{\pi}{2}~,
 \end{eqnarray}
for the case (iii), we obtain
 \begin{eqnarray}
  v^{2}_{\chi}=-\frac{1}{2}\frac{m^{2}_{\chi}}{\tilde{\lambda}^{\chi}_{1}-2(\lambda^{\chi}_{1}+\lambda^{\chi}_{2})}~,\qquad\qquad {\rm for}~\xi=\pm\frac{\pi}{4}~,
 \end{eqnarray}
 leading to
 \begin{eqnarray}
  V_{0}=-\frac{m^{4}_{\chi}}{4\tilde{\lambda}^{\chi}_{1}-8(\lambda^{\chi}_{1}+\lambda^{\chi}_{2})}~.
 \label{mini}
 \end{eqnarray}
The third case corresponds to the absolute minimum of the potential.
It could be also guaranteed that we are at a minimum by showing the eigenvalues of the neutral Higgs boson mass matrices and requiring that they are all positive.

Second, the vacuum configuration for $\Phi$ and $\eta$ are obtained by vanishing of the derivative of $V$ with respect to each component of the scalar fields $\Phi_j$ and $\eta$. The vacuum alignment of the fields $\Phi$ and $\eta$ are determined by
 \begin{eqnarray}
  \frac{\partial V}{\partial \Phi^{0}_{j}}\Big|_{\langle\Phi^{0}_j\rangle=v_{\Phi}}&=&3v_{\Phi}\left\{\mu^{2}_{\Phi}+v^{2}_{\Phi}\left(3\lambda^{\Phi}_{1}+4\lambda^{\Phi}_{3}\right)+\frac{1}{2}v^{2}_{\eta}(\lambda^{\eta\Phi}_{1}+2\lambda^{\eta\Phi}_{2}\cos2\varphi)\right\}=0~, \label{vevPhi}\\
  \frac{\partial V}{\partial \eta^{0}}\Big|_{\langle\eta^{0}\rangle=v_{\eta}}&=&v_{\eta}\left\{v^{2}_{\eta}\lambda^{\eta}+\mu^{2}_{\eta}+\frac{3}{2}v^{2}_{\Phi}(\lambda^{\eta\Phi}_{1}+2\lambda^{\eta\Phi}_{2}\cos2\varphi)\right\}=0~,
  \label{vevEta}
 \end{eqnarray}
where $j=1-3$. At the same time, with the above vacuum alignment of $\Phi$ and $\eta$ fields, the minimal condition with respect to $\phi_{i}$ and $\varphi$ are given as
 \begin{eqnarray}
 \frac{\partial V}{\partial \phi_i}\Big|_{\phi_{1}=\phi_{2}=\phi_{3}}= 0~,\qquad\frac{\partial V}{\partial \varphi}\Big|=-3v^{2}_{\eta}v^{2}_{\Phi}\lambda^{\eta\Phi}_{2}\sin2\varphi=0~,
 \end{eqnarray}
where, without loss of generality, we have let $\phi_{i}=0$ due to the interaction term $[(\Phi^{\dag}\eta)(\Phi^{\dag}\eta)]_{\mathbf{1}}$ between $\phi_{i}$ and $\varphi$.
So, we find a nontrivial VEV configuration for $\Phi$ and $\eta$ fields
 \begin{eqnarray}
  \langle\Phi\rangle=\frac{v_{\Phi}}{\sqrt{2}}(1,1,1)~,\qquad\langle\eta\rangle=\frac{v_{\eta}}{\sqrt{2}}e^{i\varphi}~.
  \label{Alin2}
 \end{eqnarray}
And, for this vacuum alignments the scalar potential can be written as
 \begin{eqnarray}
 V_{\eta\Phi}=\frac{1}{4}\left\{3v^{4}_{\Phi}\left(3\lambda^{\Phi}_{1}+4\lambda^{\Phi}_{3}\right)+6v^{2}_{\Phi}\mu^{2}_{\Phi}+v^{4}_{\eta}\lambda^{\eta}+2v^{2}_{\eta}\mu^{2}_{\eta}+3v^{2}_{\Phi}v^{2}_{\eta}(\lambda^{\eta\Phi}_{1}+2\lambda^{\eta\Phi}_{2}\cos2\varphi)\right\}~.
 \end{eqnarray}
Then, the real valued solutions are given as
 \begin{eqnarray}
  v^{2}_{\Phi}=-\frac{\mu^{2}_{\Phi}+\frac{1}{2}v^{2}_{\eta}(\lambda^{\eta\Phi}_{1}\pm2\lambda^{\eta\Phi}_{2})}{3\lambda^{\Phi}_{1}+4\lambda^{\Phi}_{3}}~,\quad v^{2}_{\eta}=-\frac{\mu^{2}_{\eta}+\frac{3}{2}v^{2}_{\Phi}(\lambda^{\eta\Phi}_{1}\pm2\lambda^{\eta\Phi}_{2})}{\lambda^{\eta}}~,
  \label{VPP1}
 \end{eqnarray}
where the plus (minus) sign in the bracket corresponds to $\varphi=0,\pm\pi$ ($\varphi=\pm\pi/2$). Those vacuum alignments do not violate CP (see later).
The VEV alignment of $\Phi$ field breaks $A_{4}$ down to a residual $Z_{3}$.

\section{Complex CKM and PMNS matrices from a common phase}
Since CP invariance has been imposed at Lagrangian level, all the parameters in the Lagrangian [see Eqs.~(\ref{poten01}), (\ref{lagrangian}) and (\ref{lagrangianCh})] are assumed to be real. We spontaneously break the $A_4$ flavor symmetry by giving nonzero complex vacuum expectation values to some components of both the $A_4$-triplets $\chi$ and $\Phi$ and the $A_4$-singlet $\eta$,
as seen in Eqs.~(\ref{Alin}) and (\ref{Alin2}).
The SM VEV $v_{\rm EW}=(\sqrt{2}G_{F})^{-1/2}=246$ GeV results from the combination $v_{\rm EW}=\sqrt{v^{2}_{\eta}+3v^{2}_{\Phi}}$. In our scenario, we assume that $v_{\chi}$ (seesaw scale) is much larger than $v_{\Phi}$ (electroweak scale):
 \begin{eqnarray}
  v_{\chi}=\lambda~\Lambda\gg v_{\Phi} \approx v_{\eta}~,
 \label{vevhier}
 \end{eqnarray}
where $\lambda$ and $\Lambda$ indicate the Cabbibo angle and the cutoff scale, respectively.

After the breaking of the flavor and electroweak symmetries, with the VEV alignments as in Eqs.~(\ref{Alin}) and (\ref{Alin2}), the charged lepton, Dirac neutrino, and right-handed neutrino mass terms from the Lagrangian~(\ref{lagrangian}) result in
 \begin{eqnarray}
 -{\cal L}_{m} &=& \frac{v_{\eta}e^{i\varphi}}{\sqrt{2}}\left(y_{e}\bar{e}_{L}e_{R}+y_{\mu}\bar{\mu}_{L}\mu_{R}+y_{\tau}\bar{\tau}_{L}\tau_{R}\right)+\frac{v_{\Phi}}{\sqrt{2}}\Big\{\left(y^{\nu}_{1}\bar{\nu}_{e}+y^{\nu}_{2}\bar{\nu}_{\mu}+y^{\nu}_{3}\bar{\nu}_{\tau}\right)N_{R1}\nonumber\\
 &+&\left(y^{\nu}_{1}\bar{\nu}_{e}+y^{\nu}_{2}\omega^{2}\bar{\nu}_{\mu}+y^{\nu}_{3}\omega\bar{\nu}_{\tau}\right)N_{R2}+\left(y^{\nu}_{1}\bar{\nu}_{e}+y^{\nu}_{2}\omega\bar{\nu}_{\mu}+y^{\nu}_{3}\omega^{2}\bar{\nu}_{\tau}\right)N_{R3}\Big\}\nonumber\\
 &+&\frac{M}{2}(\overline{N^{c}_{R1}}N_{R1}+\overline{N^{c}_{R2}}N_{R2}+\overline{N^{c}_{R3}}N_{R3})
 +\frac{y^\nu_R v_{\chi}e^{i\xi}}{2}(\overline{N^{c}_{R2}}N_{R3}+\overline{N^{c}_{R3}}N_{R2})+\text{h.c.}~.
 \label{lagrangian1}
 \end{eqnarray}
This form shows clearly that the terms in $v_\Phi$ break the $S$-flavor parity symmetry, while the other mass terms preserve it.
Inspection of the above mass terms in Eq.~(\ref{lagrangian1}) indicates that, with the VEV alignments in Eqs.~(\ref{Alin}) and (\ref{Alin2}), the $A_{4}$ symmetry is spontaneously broken to a residual $Z_{2}$ symmetry in the heavy Majorana neutrino sector (conservation of $S$-flavor parity in terms not involving $v_\Phi$) and a residual $Z_{3}$ symmetry in the Dirac neutrino sector (conservation of $T$-flavor in terms not involving $v_\chi$).

In the quark sector from the Lagrangian~(\ref{lagrangianCh}), after the breaking of the flavor and electroweak symmetries, with the VEV alignments as in Eqs.~(\ref{Alin}) and (\ref{Alin2}) the up-type quark and down-type quark mass terms result in
 \begin{eqnarray}
 {\cal L}^{q}_{m} &=&{\cal L}^{u}_{m}+{\cal L}^{d}_{m}~,
 \label{lagrangianQu1}
 \end{eqnarray}
where
 \begin{eqnarray}
-{\cal L}^{u}_{m} &=& y_{u}\frac{v_{\Phi}}{\sqrt{2}}\{\bar{u}_{L}u_{R}+\bar{c}_{L}u_{R}+\bar{t}_{L}u_{R}\}+y_{c}\frac{v_{\Phi}}{\sqrt{2}}\{\bar{u}_{L}c_{R}+\omega\bar{c}_{L}c_{R}+\omega^{2}\bar{t}_{L}c_{R}\} ~\nonumber\\
&+&y_{t}\frac{v_{\Phi}}{\sqrt{2}}\{\bar{u}_{L}t_{R}+\omega^{2}\bar{c}_{L}t_{R}+\omega\bar{t}_{L}t_{R}\}+\frac{v_{\Phi}}{\sqrt{2}}\frac{v_{\chi}e^{i\xi}}{\Lambda}\Big\{(y^{s}_{u}+iy^{a}_{u})\bar{c}_{L}u_{R}
+(y^{s}_{u}-iy^{a}_{u})\bar{t}_{L}u_{R}\nonumber\\
&+&(y^{s}_{c}+iy^{a}_{c})\bar{c}_{L}c_{R}+(y^{s}_{c}-iy^{a}_{c})\bar{t}_{L}c_{R}+(y^{s}_{t}+iy^{a}_{t})\bar{c}_{L}t_{R}+(y^{s}_{t}-iy^{a}_{t})\bar{t}_{L}t_{R}\Big\}
+\text{h.c.}~,\label{lagrangianUP}\\
-{\cal L}^{d}_{m} &=& y_{d}\frac{v_{\Phi}}{\sqrt{2}}\{\bar{d}_{L}d_{R}+\bar{s}_{L}d_{R}+\bar{b}_{L}d_{R}\}+y_{s}\frac{v_{\Phi}}{\sqrt{2}}\{\bar{d}_{L}s_{R}+\omega\bar{s}_{L}s_{R}+\omega^{2}\bar{b}_{L}s_{R}\} ~\nonumber\\
&+&y_{b}\frac{v_{\Phi}}{\sqrt{2}}\{\bar{d}_{L}b_{R}+\omega^{2}\bar{s}_{L}b_{R}+\omega\bar{b}_{L}b_{R}\}+\frac{v_{\Phi}}{\sqrt{2}}\frac{v_{\chi}e^{i\xi}}{\Lambda}\Big\{(y^{s}_{d}+iy^{a}_{d})\bar{s}_{L}d_{R}
+(y^{s}_{d}-iy^{a}_{d})\bar{b}_{L}d_{R}\nonumber\\
&+&(y^{s}_{s}+iy^{a}_{s})\bar{s}_{L}s_{R}+(y^{s}_{s}-iy^{a}_{s})\bar{b}_{L}s_{R}+(y^{s}_{b}+iy^{a}_{b})\bar{s}_{L}b_{R}+(y^{s}_{b}-iy^{a}_{b})\bar{b}_{L}b_{R}\Big\}
+\text{h.c.}~,\label{lagrangianDown}
 \end{eqnarray}
where the parameters $y^{s}_{u(c,t)}$, $y^{a}_{u(c,t)}$, $y^{s}_{d(s,b)}$ and $y^{a}_{d(s,b)}$ are all real and positive.
By taking the equal VEV alignment of $\langle \chi \rangle$ given in Eq.~(\ref{Alin}) with the
VEV alignment of $\langle \Phi \rangle$ in  Eq.~(\ref{Alin2}), the $A_{4}$ symmetry is  spontaneously broken and at the
same time its subsymmetry $Z_{3}$ is also broken through the dimension-5 operators. Including 5-dimensional operators to $V(\chi)$, the corrections to the VEV are shifted and redefined into
 \begin{eqnarray}
  \langle\chi\rangle=v_{\chi}e^{i\xi}(1+\delta_{\chi},0,0)\rightarrow v_{\chi}e^{i\xi}(1,0,0)~,
  \label{Alinchi}
 \end{eqnarray}
where the correction $\delta_{\chi}$ is dimensionless.

The nonzero expectation value $\langle \eta \rangle = v_\eta e^{i\varphi}/\sqrt{2}$ does not break the $A_4$ symmetry, because the standard model Higgs is $A_4$-flavorless.
The nonzero expectation value $\langle \Phi \rangle =v_\Phi( 1, 1, 1)/\sqrt{2}$ breaks the $S$-flavor parity but leaves the vacuum $T$-flavor $T_f=0$. In other words, after $\Phi$ acquires a nonzero VEV, the $T$-flavor is still conserved but the $S$-flavor parity is not. Since $\Phi$ appears only in the Higgs sector and in interactions with the light leptons, we say that the light neutrino sector has a residual $Z_3$ symmetry expressed by the subgroup $\{ 1, T, T^2 \}$ that leads to the conservation of $T$-flavor in terms involving mixing with the light neutrinos or interactions with the charged leptons.
The nonzero expectation value $\langle \chi \rangle = v_\chi e^{i\xi}(1,0,0)$ maintains the $S$-flavor parity of the vacuum (it is $S$-flavor-even) but gives the vacuum the symmetric combination of $T$-flavors $(a_0+a_{+1}+a_{-1})/\sqrt{3}$. That is, after $\chi$ acquires a nonzero VEV, the $S$-flavor parity is  conserved but the $T$-flavor is not.  Since $\chi$ appears only in the Higgs sector and in interactions with the heavy Majorana neutrinos, we say that the heavy neutrino sector has a residual $Z_2$ symmetry expressed by the subgroup $\{1,S\}$ leading to the conservation of  $S$-flavor parity in terms involving mixing or interactions with the heavy Majorana neutrinos.

\subsection{Quark sector and CKM matrix}
With the help of the VEVs of the $A_{4}$-triplet $\Phi$ which is equally aligned, that is,
$\langle \Phi \rangle = (1,1,1)v_{\Phi}/\sqrt{2}$ in Eq.~(\ref{Alin2}), the up-type quark mass matrix $\mathcal{M}_{u}$ can be explicitly expressed as
 \begin{eqnarray}
 \mathcal{M}_{u}&=&U_{\omega}\sqrt{\frac{3}{2}}{\left(\begin{array}{ccc}
 y_{u} & 0 & 0 \\
 0 & y_{c} & 0 \\
 0 & 0 & y_{t}
 \end{array}\right)}v_{\Phi}-U_{\omega}\sqrt{\frac{3}{2}}{\left(\begin{array}{ccc}
 -\frac{2y^s_{u}}{3} & -\frac{2y^s_{c}}{3} & -\frac{2y^s_{t}}{3} \\
 \frac{y^s_{u}-\sqrt{3}y^{a}_{u}}{3} & \frac{y^s_{c}-\sqrt{3}y^{a}_{c}}{3} & \frac{y^s_{t}-\sqrt{3}y^{a}_{t}}{3} \\
 \frac{y^s_{u}+\sqrt{3}y^{a}_{u}}{3} & \frac{y^s_{c}+\sqrt{3}y^{a}_{c}}{3} & \frac{y^s_{t}+\sqrt{3}y^{a}_{t}}{3}
 \end{array}\right)}v_{\Phi} e^{i\xi}\frac{v_{\chi}}{\Lambda}~\nonumber\\
 &=& U_{\omega}V^{u}_{L} ~{\rm Diag}(m_{u},m_{c},m_{t})~ V^{u\dag}_{R}~,
 \label{UPcorrect}
 \end{eqnarray}
where $U_{\omega}V^{u}_{L}$ and $V^{u}_{R}$ are the diagonalization matrices for $\mathcal{M}_{u}$,
and
\begin{eqnarray}
 U_{\omega}=\frac{1}{\sqrt{3}}{\left(\begin{array}{ccc}
 1 &  1 &  1 \\
 1 &  \omega &  \omega^{2} \\
 1 &  \omega^{2} &  \omega
 \end{array}\right)}~.
 \label{Uomega}
 \end{eqnarray}
And the down-type quark mass matrix $\mathcal{M}_{d}$ can be explicitly expressed as
 \begin{eqnarray}
 \mathcal{M}_{d}&=&U_{\omega}\sqrt{\frac{3}{2}}{\left(\begin{array}{ccc}
 y_{d} & 0 & 0 \\
 0 & y_{s} & 0 \\
 0 & 0 & y_{b}
 \end{array}\right)}v_{\Phi}-U_{\omega}\sqrt{\frac{3}{2}}{\left(\begin{array}{ccc}
 -\frac{2y^s_{d}}{3} & -\frac{2y^s_{s}}{3} & -\frac{2y^s_{b}}{3} \\
 \frac{y^s_{d}-\sqrt{3}y^{a}_{d}}{3} & \frac{y^s_{s}-\sqrt{3}y^{a}_{s}}{3} & \frac{y^s_{b}-\sqrt{3}y^{a}_{b}}{3} \\
 \frac{y^s_{d}+\sqrt{3}y^{a}_{d}}{3} & \frac{y^s_{s}+\sqrt{3}y^{a}_{s}}{3} & \frac{y^s_{b}+\sqrt{3}y^{a}_{b}}{3}
 \end{array}\right)}v_{\Phi} e^{i\xi}\frac{v_{\chi}}{\Lambda}~\nonumber\\
 &=& U_{\omega}V^{d}_{L} ~{\rm Diag}(m_{d},m_{s},m_{b})~ V^{d\dag}_{R}~,
 \label{Downcorrect}
 \end{eqnarray}
where $U_{\omega}V^{d}_{L}$ and $V^{d}_{R}$ are the diagonalization matrices for $\mathcal{M}_{d}$.

One of the most interesting features observed by experiments on the quarks is that the mass
spectra are strongly hierarchical, {\it i.e.}, the masses of the third
generation quarks are much heavier than those of the first and second generation quarks.
For the elements of $\mathcal{M}_{u(d)}$ given in Eqs.~(\ref{UPcorrect}) and (\ref{Downcorrect}), taking into account the most natural case
that the quark Yukawa couplings have the strong hierarchy ~$y_{f_{3}} \gg y_{f_{2}}
\gg y_{f_{1}}$~(here $f_{i}$ stands for $i$-th generation of $f$-type quark) and the off-diagonal elements generated by the higher dimensional operators are
generally smaller in magnitude than the diagonal ones, we make a plausible assumption
 \begin{eqnarray}
  y_{f_{i}} &\gg& \frac{v_{\chi}}{\Lambda}|y^f_{2i}|\sim ~({\rm or}\gg)~ \frac{v_{\chi}}{\Lambda}|y^f_{3i}|\gg \frac{v_{\chi}}{\Lambda}|y^f_{1i}| ~,
 \label{hierarchy}
 \end{eqnarray}
where $y^f_{ji}$ stands for the $ji$-component of an $f$-type quark.
Then $V^{f}_{L}$ and $V^{f}_{R}$ can be determined by diagonalizing the matrices
$\mathcal{M}_{f}\mathcal{M}^{\dag}_{f}$ and $\mathcal{M}^{\dag}_{f}\mathcal{M}_{f}$, respectively, indicated from
Eqs.~(\ref{UPcorrect}) and (\ref{Downcorrect}). Especially, the mixing matrix $V^{f}_{L}$ becomes one of the matrices composing
the CKM ones and it can be approximated, due to the
strong hierarchy expressed in Eq.~(\ref{hierarchy}), as~\cite{Ahn:2011yj}
 \begin{eqnarray}
 V^{f}_{L}=U^f_{L}Q_f
 \label{mixingL}
 \end{eqnarray}
where a diagonal phase matrix $Q_f={\rm diag}(e^{i\xi^{f}_{1}},e^{i\xi^{f}_{2}},e^{i\xi^{f}_{3}})$, which can be rotated away by the redefinition of left-handed quark fields, and
 \begin{eqnarray}
 \footnotesize
 U^{f}_{L}\simeq
 {\left(\begin{array}{ccc}
 1 - \frac{1}{2} \left| \frac{\mathcal{M}^{f}_{12}}{\mathcal{M}^{f}_{22}} \right|^{2}
  & \left| \frac{\mathcal{M}^{f}_{12}}{\mathcal{M}^{f}_{22}} \right| e^{i\phi^{f}_{3}}
  & \left| \frac{\mathcal{M}^{f}_{13}}{\mathcal{M}^{f}_{33}} \right| e^{i\phi^{f}_{2}}  \\
 - \left| \frac{\mathcal{M}^{f}_{12}}{\mathcal{M}^{f}_{22}} \right| e^{-i\phi^{f}_{3}}
  & 1 -\frac{1}{2} \left| \frac{\mathcal{M}^{f}_{12}}{\mathcal{M}^{f}_{22}} \right|^{2}
  & \left| \frac{\mathcal{M}^{f}_{23}}{\mathcal{M}^{f}_{33}} \right| e^{i\phi^{f}_{1}} \\
 - \left| \frac{\mathcal{M}^{f}_{13}}{\mathcal{M}^{f}_{33}} \right| e^{-i\phi^{f}_{2}}
   + \left| \frac{\mathcal{M}^{f}_{12}}{\mathcal{M}^{f}_{22}} \right|
   \left| \frac{\mathcal{M}^{f}_{23}}{\mathcal{M}^{f}_{33}} \right| e^{-i(\phi^{f}_{3}+\phi^{f}_{1})}
  & - \left| \frac{\mathcal{M}^{f}_{23}}{\mathcal{M}^{f}_{33}} \right| e^{-i\phi^{f}_{1}}
   - \left| \frac{\mathcal{M}^{f}_{13}}{\mathcal{M}^{f}_{33}} \right|
    \left| \frac{\mathcal{M}^{f}_{12}}{\mathcal{M}^{f}_{22}} \right| e^{i(\phi^{f}_{3}-\phi^{f}_{2})}
  & 1
 \end{array}\right)}.\nonumber
 \end{eqnarray}

There exist several empirical fermion mass ratios in the up- and down-type quark sectors
calculated from the measured values~\cite{PDG} :
 \begin{eqnarray}
  \frac{m_{d}}{m_{b}} &\simeq&  1.2\times 10^{-3}~,\quad
  \frac{m_{s}}{m_{b}} \simeq 2.4\times10^{-2}~,\quad
  \frac{m_{u}}{m_{t}} \simeq 1.4\times 10^{-5}~,\quad
  \frac{m_{c}}{m_{t}}\simeq7.4\times10^{-3}~,
 \label{massRatio}
 \end{eqnarray}
which shows that the mass spectrum of the up-type quarks exhibits a much stronger hierarchical pattern to that of the down-type quarks.
In terms of the Cabbibo angle $\lambda \equiv \sin\theta_{\rm C} \approx |V_{us}|$, the
quark masses scale as~$(m_{d},m_{s}) \approx (\lambda^{4},\lambda^{2})~ m_{b}$~ and
~$(m_{u},m_{c}) \approx (\lambda^{8},\lambda^{4})~ m_{t}$,~ which may represent the following
fact: the CKM matrix is mainly generated by the mixing matrix of the down-type quark sector,
when the Lagrangian~(\ref{lagrangianCh}) is also taken into account.

\subsubsection{The up-type quark sector and its mixing matrix}

From Eq.~(\ref{UPcorrect}) we see that the up-type quark mass matrix $\mathcal{M}_{u}$ can be diagonalized
in the mass basis by a biunitary transformation,
$V^{u\dag}_{L}\mathcal{M}_{u}V^{u}_{R}={\rm Diag}(m_{u},m_{c},m_{t})$.
The matrices $V^{u}_{L}$ and $V^{u}_{R}$ can be determined by diagonalizing the matrices
$\mathcal{M}_{u}\mathcal{M}^{\dag}_{u}$ and $\mathcal{M}^{\dag}_{u}\mathcal{M}_{u}$, respectively.
Especially, the left-handed up-type quark mixing matrix $V^{u}_{L}$ becomes one of the matrices
composing the CKM matrix such as $V_{\rm CKM} \equiv V^{u\dag}_{L}V^{d}_{L}$ (see Eq.~(\ref{ckm1})
below).
Due to the measured value of $m_u / m_t$ in Eq.~(\ref{massRatio}), it is impossible to generate the
Cabbibo angle, $\lambda\approx|V_{us}|$, from the mixing between the first and second generations in
the up-type quark sector: if one sets ~$|(V^{u}_{L})_{12}| = |\mathcal{M}^{u}_{12}/\mathcal{M}^{u}_{22}|\approx\lambda$,~
then from Eq.~(\ref{hierarchy}) one obtains
~$m_{u}/m_{t}\approx|\mathcal{M}^{u}_{12}/\mathcal{M}^{u}_{22}|~ |\mathcal{M}^{u}_{22}/\mathcal{M}^{u}_{33}|\approx\lambda^{5}$,~ in
discrepancy with the measured $m_{u}/m_{t}\approx\lambda^{8}$ in Eq.~(\ref{massRatio}).
To determine the correct up-type quark mixing matrix, using both Eqs.~(\ref{hierarchy})
and (\ref{massRatio}), we obtain $m_{c}/m_{t}\approx|\mathcal{M}^{u}_{22}/\mathcal{M}^{u}_{33}|\approx \lambda^{4}$,
$m_{u}/m_{c}\approx|\mathcal{M}^{u}_{11}/\mathcal{M}^{u}_{22}|\approx \lambda^{4}$ and
$m_{u}/m_{t}\approx|\mathcal{M}^{u}_{11}/\mathcal{M}^{u}_{33}|\approx \lambda^{8}$.
The above can be realized in our model through
 \begin{eqnarray}
 \frac{2y^{s}_{c}}{3y_{c}}~,~\frac{2y^{s}_{t}}{3y_{t}}~,~\frac{y^{s}_{t}-\sqrt{3}y^{a}_{t}}{3y_{t}}\lesssim{\cal O}(\lambda^{3})~.
 \end{eqnarray}
In particular, for a case normalized by the top quark mass :
 \begin{eqnarray}
 1\gg \frac{y^{s}_{t},y^{a}_{t}}{y_{t}}\sim{\cal O}(\lambda^{3})\gg \frac{y_{c}}{y_{t}}\sim{\cal O}(\lambda^{4})\gg\frac{y^{s}_{c},y^{a}_{c}}{y_{t}}\sim{\cal O}(\lambda^{7})\gg\frac{y_{u}}{y_{t}}\sim{\cal O}(\lambda^{8})\gg \frac{y^{s}_{u},y^{a}_{u}}{y_{t}}~,
 \label{hierarchyU}
 \end{eqnarray}
under the constraint of unitarity, the up-type quark mixing matrix $V^u_L$ can be approximated as
 \begin{eqnarray}
 V^{u}_{L}\simeq
{\left(\begin{array}{ccc}
 1 & \lambda^{4}e^{i\phi^{u}_{3}}  & \lambda^{4}e^{i\phi^{u}_{2}}  \\
 - \lambda^{4}e^{-i\phi^{u}_{3}} &  1 & \lambda^{4}e^{i\phi^{u}_{1}} \\
 - \lambda^{4}e^{-i\phi^{u}_{2}} &  - \lambda^{4}e^{-i\phi^{u}_{1}} & 1
 \end{array}\right)}Q_{u} +{\cal O}(\lambda^{5}) ~,
 \label{UL}
 \end{eqnarray}
which indicates that the mixing in the up-type quark sector does not affect the leading order
contributions in $\lambda$.  It leads to the fact that the Cabbibo angle should arise from the mixing
between the first and second generations in the down-type quark sector.

\subsubsection{The down-type quark sector and its mixing matrix}

Now let us consider the down-type quark sector
to obtain the realistic CKM matrix. From Eq.~(\ref{hierarchy}) and the measured down-type quark mass
hierarchy in Eq.~(\ref{massRatio}), we find
~$m_{s}/m_{b}\approx|\mathcal{M}^{d}_{22}/\mathcal{M}^{d}_{33}| \approx 0.6 \,\lambda^{2}$,
~$m_{d}/m_{b}\approx|\mathcal{M}^{d}_{11}/\mathcal{M}^{d}_{33}| \approx 0.7 \,\lambda^{4}$~ and
~$m_{d}/m_{s}\approx|\mathcal{M}^{d}_{11}/\mathcal{M}^{d}_{22}| \approx \lambda^{2}$.~
From Eqs.~(\ref{hierarchy}) and (\ref{mixingL}), we obtain
$|(V^{d}_{L})_{12}| \approx |\mathcal{M}^{d}_{12}/\mathcal{M}^{d}_{22}| \approx 1.7 \,\lambda^{-2} |\mathcal{M}^{d}_{12}/\mathcal{M}^{d}_{33}|$,
which means ~$|\mathcal{M}^{d}_{12}/\mathcal{M}^{d}_{33}| \approx 0.6 \,\lambda^{3}$~ for $|(V^{d}_{L})_{12}|
\approx \lambda$.~  In order to get the correct CKM matrix element
$|\mathcal{M}^{d}_{13} / \mathcal{M}^{d}_{33}| \sim {\cal O}(\lambda^{3})$, we need to make an additional assumption:
from Eq.~(\ref{hierarchy}) the hierarchy normalized by the bottom quark mass can be
expressed as
 \begin{eqnarray}
 1\gg \frac{y^{a}_{b}}{y_{b}}\sim{\cal O}(\lambda)\gg \frac{y^{s}_{b}}{y_{b}}\sim\frac{y_{s}}{y_{b}}\sim\frac{y^{a}_{s},y^{s}_{s}}{y_{b}}\sim{\cal O}(\lambda^{2})\gg\frac{y_{d}}{y_{b}}\sim{\cal O}(\lambda^{4})\gg \frac{y^{a}_{d},y^{s}_{d}}{y_{b}}\sim{\cal O}(\lambda^{5})~.
 \label{hierarchyD}
 \end{eqnarray}
Then, we can obtain the mixing elements in $V^d_L$ of the down-type quarks, in a good approximation, as
 \begin{eqnarray}
  \quad\left| \frac{\mathcal{M}^{d}_{23}}{\mathcal{M}^{d}_{33}} \right|&\simeq&\lambda\frac{y^{a}_{b}}{\sqrt{3}y_{b}}~,\quad\qquad \phi^{d}_{1}\simeq\frac{1}{2}\arg\left\{-e^{i\xi}\left(\frac{y^{s}_{b}}{y_{b}}-\sqrt{3}\frac{y^{a}_{b}}{y_{b}}\right)\right\}~,\nonumber\\
  \left| \frac{\mathcal{M}^{d}_{13}}{\mathcal{M}^{d}_{33}} \right|&\simeq&\lambda\frac{2y^{s}_{b}}{3y_{b}}~,~~\quad\qquad \phi^{d}_{2}\simeq\frac{1}{2}\arg\left\{e^{i\xi}-\frac{\lambda}{3}\left(\frac{y^{s}_{b}}{y_{b}}-\sqrt{3}\frac{y^{a}_{b}}{y_{b}}\right)\right\}~,\nonumber\\
  \left| \frac{\mathcal{M}^{d}_{12}}{\mathcal{M}^{d}_{22}} \right|&\simeq&\lambda\frac{2y^{s}_{s}}{3y_{s}}~,~~\qquad\quad\phi^{d}_{3}\simeq\frac{1}{2}\arg\left\{e^{i\xi}-\frac{\lambda}{3}\left(\frac{y^{s}_{s}}{y_{s}}+\sqrt{3}\frac{y^{a}_{s}}{y_{s}}\right)\right\}~.
 \label{mixingD}
 \end{eqnarray}
Here, the phase $\phi^{d}_{i}$ ($i=1,2,3$) mainly depends on the parameter $\xi$.
Under the constraint of unitarity,
the mixing matrix $V^d_L$ can be written in terms of Eqs.~(\ref{hierarchyD}) and (\ref{mixingD}) as
 \begin{eqnarray}
 V^{d}_{L}\simeq
 {\left(\begin{array}{ccc}
 1 -\frac{\lambda^{2}}{2} & \lambda e^{i\phi^{d}_{3}}  & A' \lambda^{3} e^{i\phi^{d}_{2}}  \\
 - \lambda e^{-i\phi^{d}_{3}} &  1-\frac{\lambda^{2}}{2} & A \lambda^{2} e^{i\phi^{d}_{1}} \\
 -A' \lambda^{3} e^{-i\phi^{d}_{2}} +A \lambda^{3} e^{-i(\phi^{d}_{3}+\phi^{d}_{1})}
  &  -A \lambda^{2} e^{-i\phi^{d}_{1}} & 1
 \end{array}\right)}Q_{d}+{\cal O}(\lambda^{4})~,
 \label{DL}
 \end{eqnarray}
where we have used the following:
 \begin{eqnarray}
 \frac{y^{a}_{b}}{\sqrt{3}y_{b}}=A\lambda~,\qquad\frac{2y^{s}_{b}}{3y_{b}}=A'\lambda^{2}~,\qquad2y^{s}_{s}=3y_{s}~.
 \label{DL1}
 \end{eqnarray}
Later in Eq.~(\ref{ckm1}), we shall see
that this form of $V^d_L$ indeed becomes the realistic CKM matrix. And the mass squared eigenvalues are written in terms of Eq.~(\ref{DL1}) as
 \begin{eqnarray}
 m^{2}_{d}&\simeq&\frac{v^{2}_{\Phi}}{2}y^{2}_{d}\left\{1+\frac{4y^{s}_{d}}{3y_{d}}\lambda\cos\xi\right\}~,\nonumber\\
 m^{2}_{s}&\simeq&\frac{v^{2}_{\Phi}}{2}y^{2}_{s}\left\{1-\lambda\cos\xi\left(1+\frac{y^{a}_{s}}{y_{s}}\right)\right\}~,\nonumber\\
  m^{2}_{b}&\simeq&\frac{v^{2}_{\Phi}}{2}y^{2}_{b}\left\{1+\lambda^{2}\cos\xi(A\sqrt{3}-\lambda A')\right\}~.
 \label{Dmass}
 \end{eqnarray}

\subsubsection{CKM mixing matrix}
In the weak eigenstate basis, the quark mass terms in Eq.~(\ref{lagrangianQu1}) and the charged gauge
interactions can be written as
 \begin{eqnarray}
 -{\cal L}_{q W} &=& \overline{q^{u}_{L}}\mathcal{M}_{u}q^{u}_{R}+\overline{q^{d}_{L}}\mathcal{M}_{d}q^{d}_{R}
  +\frac{g}{\sqrt{2}}W^{+}_{\mu} ~\overline{q^{u}_{L}}\gamma^{\mu}q^{d}_{L} + {\rm h.c.} ~.
 \label{lagrangianA}
 \end{eqnarray}
From Eq.~(\ref{lagrangianA}), to diagonalize the charged fermion mass matrices such
that
 \begin{eqnarray}
  V^{f\dag}_{L}~\mathcal{M}_{f}~ V^{f}_{R} = {\rm Diag}(m_{f_{1}},m_{f_{2}},m_{f_{3}})~
  \equiv \hat m_f  ~,
 \end{eqnarray}
we can rotate the fermion fields from the weak eigenstates to the mass eigenstates:
 \begin{eqnarray}
   q^{u(d)}_{L} \rightarrow V^{u(d) \dag}_{L}~ q^{u(d)}_{L} ~,\qquad
   q^{u(d)}_{R} \rightarrow V^{u(d)\dag}_{R} q^{u(d)}_{R} ~.
\label{basis}
 \end{eqnarray}
Then, from the charged current terms in Eq.~(\ref{lagrangianA}), we obtain the CKM matrix
 \begin{eqnarray}
  V_{\rm CKM} &=& V^{u\dag}_{L}V^{d}_{L}~.
 \label{CKMmixing}
 \end{eqnarray}
From Eqs.~(\ref{UL}) and (\ref{DL}), with the transformations $s_{L} \to s_{L}~e^{-i\phi^{d}_{3}}$,
$b_{L} \to b_{L}~e^{-i(\phi^{d}_{1}+\phi^{d}_{3})}$, $c_{L} \to c_{L}~e^{-i\phi^{d}_{3}}$ and
$t_{L} \to t_{L}~e^{-i(\phi^{d}_{1}+\phi^{d}_{3})}$, if we set
 \begin{eqnarray}
  A' e^{-i(\phi^{d}_{1}+\phi^{d}_{3}-\phi^{d}_{2})} = A (\rho -i\eta) ~,
 \end{eqnarray}
then we obtain the CKM matrix in the Wolfenstein parametrization~\cite{Wolfenstein:1983yz} given by
 \begin{eqnarray}
  V_{\rm CKM}=V^{u\dag}_{L}V^{d}_{L}\simeq V^{d}_{L}
  \simeq{\left(\begin{array}{ccc}
 1-\frac{\lambda^{2}}{2} & \lambda  & A\lambda^{3}(\rho-i\eta)  \\
 -\lambda &  1-\frac{\lambda^{2}}{2} & A\lambda^{2} \\
 A\lambda^{3}(1-\rho-i\eta) & -A\lambda^{2}  & 1
 \end{array}\right)}+{\cal O}(\lambda^{4})~.
 \label{ckm1}
 \end{eqnarray}
As reported in Ref.~\cite{ckmfitter} the best-fit values of the parameters $\lambda$, $A$, $\bar{\rho}$,
$\bar{\eta}$ with $1\sigma$ errors are
 \begin{eqnarray}
  \lambda &=& \sin\theta_{C}=0.22543\pm0.00077~,~~~~~A=0.812^{+0.013}_{-0.027}~, \nonumber\\
  \bar{\rho} &=& 0.144\pm0.025~,~~~~~~~~~~~~~~~~~~~~~~~\bar{\eta}=0.342^{+0.016}_{-0.015}~,
 \label{CKMpara}
 \end{eqnarray}
where $\bar{\rho}=\rho(1-\lambda^{2}/2)$ and $\bar{\eta}=\eta(1-\lambda^{2}/2)$. The effects caused by
CP violation are always proportional to the Jarlskog invariant~\cite{Jarlskog:1985ht}, defined as
$J^{\rm quark}_{CP} = {\rm Im}[V_{us}V_{cb}V^{\ast}_{ub}V^{\ast}_{cs}] \simeq A^{2} \lambda^{6}\eta$
whose value is $2.96^{+0.18}_{-0.17}\times10^{-5}$ at $1\sigma$ level~\cite{ckmfitter}.
In terms of the Wolfenstein parametrization the mixing parameters $A'$ and $\phi^{d}_{i}$ can be interpreted as
 \begin{eqnarray}
  A'=A\sqrt{\rho^{2}+\eta^{2}}~,\qquad \delta^{q}_{CP}\equiv\phi^{d}_{1}+\phi^{d}_{3}-\phi^{d}_{2}=\tan^{-1}\left(\frac{\eta}{\rho}\right)~.
 \end{eqnarray}
Putting Eq.~(\ref{DL1}) and the ratio $y^{a}_{s}/y_{s}=x\sim{\cal O}(1)$ into the phases $\phi^{d}_{i}$ in Eq.~(\ref{mixingD}), we obtain
 \begin{eqnarray}
  \phi^{d}_{1}&\simeq&\frac{1}{2}\arg\{e^{i(\xi+\pi)}(\lambda-3A)\}~,\nonumber\\ \phi^{d}_{2}&\simeq&\frac{1}{2}\arg\{e^{i\xi}+A\lambda^{2}\}~,\qquad\qquad
  \phi^{d}_{3}\simeq\frac{1}{2}\arg\{e^{i\xi}-\frac{\lambda}{3}(1+x\sqrt{3})\}~.
 \label{CKMpara2}
 \end{eqnarray}
In our model the CKM Dirac CP phase explicitly depends on the phase $\xi$ associated with the leptonic Dirac CP phase: for example, taking $\xi=120^{\circ}~(110^{\circ})$ for $x=1~(4)$ we obtain $\delta^{q}_{CP}\simeq67^{\circ}$ which is in a good agreement with the present data.

\subsubsection{The strong CP problem}
There is a common problem in models with spontaneous CP violation, which is that a strong QCD $\bar{\vartheta}_{\rm eff}$ term will be generated~\cite{SCPV}.
The associated strong CP problem is written as
 \begin{eqnarray}
  \bar{\vartheta}_{\rm eff}=\vartheta+\arg\left\{\det(\mathcal{M}_{u})\det(\mathcal{M}_{d})\right\}.
 \end{eqnarray}
The $\bar{\vartheta}_{\rm eff}$ is the coefficient of
$\bar{\vartheta}_{\rm eff}~F^{a}_{\mu\nu}\tilde{F}^{\mu\nu a}/32\pi^{2}$. The second term
in the above equation comes from a chiral transformation for diagonalization of the quark mass matrices. Experimental bounds on CP violation in strong interactions are very tight, the strongest one coming from the limits on the electric dipole moment of the neutron $d_{n}<0.29\times10^{-25}~e$~\cite{Beringer:1900zz} which implies $|\bar{\vartheta}_{\rm eff}|<0.56\times10^{-10}$. $\bar{\vartheta}_{\rm eff}$ should be very small to make a theory consistent with experimental bounds.
A huge cancellation between $\vartheta$ and $\arg\left\{\det(\mathcal{M}_{u})\det(\mathcal{M}_{d})\right\}$ suggests that there should be a physical process.

At tree level the strong CP problem is automatically solved, {\it i.e.} $\bar{\vartheta}_{\rm eff}^{\rm tree}=0$ : the term $\vartheta$ vanishes since the CP symmetry is imposed at the Lagrangian level, and since the matrices $\mathcal{M}_{u(d)}$ are real diagonal [which can be achieved by the rotation of $q^{u(d)}_{L}\rightarrow U^{\dag}_{\omega}q^{u(d)}_{L}$ in Eqs.~(\ref{lagrangianUP}), (\ref{lagrangianDown}) and (\ref{lagrangianA}) at the tree level with $V_{\rm CKM}=U^{\dag}_{\omega}U_{\omega}=$ unit matrix], the term $\arg\left\{\det(\mathcal{M}_{u})\det(\mathcal{M}_{d})\right\}$ is zero.
Including higher dimensional operators, the situation is changed. If the first contribution of up-type quark to the CKM matrix appears in the order of $\lambda^4$, {\it i.e.} Eq.~(\ref{UL}), its contribution
to the  $\bar{\vartheta}_{\rm eff}$ can be estimated as
 \begin{eqnarray}
\bar{\vartheta}_{\rm eff} &=& \arg\left\{\det(\mathcal{M}_{u})\det(\mathcal{M}_{d})\right\} \nonumber\\
&\approx& \arg \left\{\det(V_L^u \hat{\mathcal{M}}_{u} V_R^u)
\det(V_L^u V_{\rm CKM} \hat{\mathcal{M}}_{d} V_R^d)\right\}
\lesssim {\cal O}(\lambda^5) \approx 10^{-4}~.
 \label{STCP}
 \end{eqnarray}
This value is well above the required $10^{-9}$ level and we may need some additional dynamical mechanism to suppress it. However, we can show
the vanishing $\bar{\vartheta}_{\rm eff}$ is consistent with our model,
although we do not solve the strong CP problem. To see this we can write
 \begin{eqnarray}
\arg\left\{\det(\mathcal{M}_{u})\det(\mathcal{M}_{d})\right\}=\arg\left\{1+\sum^{6}_{N=1}\left(\frac{v_{\chi}}{\Lambda}\right)^{N}e^{iN\xi}(...)\right\}<10^{-9-10}~,
 \label{STCP}
 \end{eqnarray}
where $``..."$ stands for combinations of $\frac{y^{a(s)}_{u}}{y_{u}},\frac{y^{a(s)}_{c}}{y_{c}},\frac{y^{a(s)}_{t}}{y_{t}},\frac{y^{a(s)}_{d}}{y_{d}},\frac{y^{a(s)}_{s}}{y_{s}}$, and $\frac{y^{a(s)}_{b}}{y_{b}}$;
one can suppress the contributions of operators with dimensions higher than 5~\footnote{Here, we do not consider the suppression of loop effects under renormalizability on the $\bar{\vartheta}_{\rm eff}$ parameter.}.
For example, we can consider a scenario where the entire CKM mixing matrix comes from the down-type quark
sector. This is legitimate because
the up-type quark mass hierarchy is much stronger than the down-type one
and as a consequence the up-type quark contribution to the CKM matrix is small as can be
seen from Eq.~(\ref{UL}).
This corresponds to neglecting the second term in Eq.~(\ref{UPcorrect}). Then
the contributions of $N\geq3$ in Eq.~(\ref{STCP}) are automatically zero.
With the choice
$\frac{y^{s}_{d}}{y_{d}}=\frac{3}{2}A\lambda(1+\frac{1}{2}\lambda\sqrt{\rho^{2}+\eta^{2}})\sim{\cal O}(\lambda)$
and $\frac{y^{a}_{d}}{y_{d}}=\frac{3\sqrt{3}}{2}A\lambda\frac{1-\frac{A}{\sqrt{3}}\lambda-\lambda\frac{A}{6}\sqrt{\rho^{2}+\eta^{2}}(1+\sqrt{3}\lambda)}{1-\frac{A}{2}\sqrt{3(\rho^{2}+\eta^{2})}\lambda^{2}}\sim{\cal O}(\lambda)$,
which obey the scaling rules in Eq.~(\ref{hierarchyD}),
we obtain $\arg\left\{\det(\mathcal{M}_{u})\det(\mathcal{M}_{d})\right\}=0$ irrespective of the phase $\xi$.

Including higher dimensional operators to the quark Yukawa Lagrangian, that is,
 \begin{eqnarray}
  &&(\bar{Q}_{L}\tilde{\Phi})_{{\bf 1}}(\chi\chi^{(\ast)})_{{\bf 1}}u_{R}~,\quad(\bar{Q}_{L}\tilde{\Phi})_{{\bf 1}'}(\chi\chi^{(\ast)})_{{\bf 1}}c_{R}~,...\nonumber\\
  &&(\bar{Q}_{L}\Phi)_{{\bf 1}}(\chi\chi^{(\ast)})_{{\bf 1}}d_{R}~,\quad(\bar{Q}_{L}\Phi)_{{\bf 1}'}(\chi\chi^{(\ast)})_{{\bf 1}}s_{R}~,...
 \end{eqnarray}
the corrections to the mass terms Eqs.~(\ref{UPcorrect}) and (\ref{Downcorrect}) are shifted just in the tree level mass terms and redefined into
 \begin{eqnarray}
 &&U_{\omega}\cdot{\rm Diag}(y_{u}+\delta_{u},y_{c}+\delta_{c},y_{t}+\delta_{t})\rightarrow U_{\omega}\cdot{\rm Diag}(y_{u},y_{c},y_{t})~,\nonumber\\
 &&U_{\omega}\cdot{\rm Diag}(y_{d}+\delta_{d},y_{s}+\delta_{s},y_{b}+\delta_{b})\rightarrow U_{\omega}\cdot{\rm Diag}(y_{d},y_{s},y_{b})~,
 \end{eqnarray}
where the correction $\delta_{f}$ ($f=u,c,t,d,s,b$) is dimensionless, and the following operators do not affect the corrections
 \begin{eqnarray}
  &&(\bar{Q}_{L}\tilde{\Phi})_{{\bf 3}}(\chi\chi^{(\ast)})_{{\bf 3}}u_{R}~,\quad[(\bar{Q}_{L}\tilde{\Phi})_{{\bf 3}}(\chi\chi^{(\ast)})_{{\bf 3}}]_{{\bf 1}'}c_{R}~,...\nonumber\\
  &&(\bar{Q}_{L}\Phi)_{{\bf 3}}(\chi\chi^{(\ast)})_{{\bf 3}}d_{R}~,\quad[(\bar{Q}_{L}\Phi)_{{\bf 3}}(\chi\chi^{(\ast)})_{{\bf 3}}]_{{\bf 1}'}s_{R}~,...
 \end{eqnarray}
due to the VEV alignment of the $\chi$ field in Eq.~(\ref{Alin}). Thus, the effects of higher dimensional ($d\geq6$) operators to the strong CP problem may be equivalent to the one of the dimension-5 operators.

\subsection{Lepton sector and PMNS matrix}
The leptonic mass terms in Eq.~(\ref{lagrangian1}) and the charged gauge interactions in the weak eigenstate basis can be written in (block) matrix form as
 \begin{eqnarray}
 -{\cal L}_{\ell W} &=& \frac{1}{2}\overline{N^{c}_{R}}M_{R}N_{R}
 +\overline{\nu_{L}}m_{D}N_{R}+\overline{\ell_{L}}\mathcal{M}_{\ell}\ell_{R}+\frac{g}{\sqrt{2}}W^{-}_{\mu}\overline{\ell_{L}}\gamma^{\mu}\nu_{L}+\text{h.c.}~ \\
 &=& \frac{1}{2} \begin{pmatrix} \overline{\nu_L} & \overline{N^{c}_R} \end{pmatrix} \begin{pmatrix} 0 & m_D \\ m_D^T & M_R \end{pmatrix} \begin{pmatrix} \nu^{c}_L \\ N_R \end{pmatrix} + \overline{\ell_{L}}\mathcal{M}_{\ell}\ell_{R}+\frac{g}{\sqrt{2}}W^{-}_{\mu}\overline{\ell_{L}}\gamma^{\mu}\nu_{L}+\text{h.c.}
 \label{lagrangianA1}
 \end{eqnarray}
Here $\ell=(e,\mu,\tau)$, $\nu=(\nu_e,\nu_\mu,\nu_\tau)$, $N_R=(N_{R1},N_{R2},N_{R3})$, and
\begin{align}
\mathcal{M}_{\ell} & =\frac{v_{\eta}e^{i\varphi}}{\sqrt{2}}
\begin{pmatrix}
y_{e} & 0 & 0 \\
0 & y_{\mu} & 0 \\
0 & 0 & y_{\tau}
\end{pmatrix} , \\
m_{D}& = \frac{v_{\Phi}}{\sqrt{2}} \begin{pmatrix}
y^{\nu}_1 & y^{\nu}_1 & y^{\nu}_1 \\
y^{\nu}_2 & \omega^2 y^{\nu}_2 & \omega y^{\nu}_2 \\
y^{\nu}_3 & \omega y^{\nu}_3 & \omega^2 y^{\nu}_3
\end{pmatrix} =v_{\Phi}y^{\nu}_{1}\sqrt{\frac{3}{2}}{\left(\begin{array}{ccc}
 1 &  0 &  0 \\
 0 & y_{2} & 0 \\
 0 & 0 & y_{3}
 \end{array}\right)}U^{\dag}_{\omega}\equiv\frac{v_{\Phi}}{\sqrt{2}}Y_{\nu},
\label{eq:Ynu}
\\
M_R & = \begin{pmatrix}
M & 0 & 0 \\
0 & M & y^{\nu}_{R} v_\chi e^{i\xi} \\
0 & y^{\nu}_{R} v_\chi e^{i\xi} & M
\end{pmatrix} ,
\end{align}
where $y_{2}=y^{\nu}_{2}/y^{\nu}_{1}, y_{3}=y^{\nu}_{3}/y^{\nu}_{1}$, and $U_\omega$ is given in Eq.~(\ref{Uomega}).

We start by diagonalizing $M_R$. For this purpose, we perform a basis rotation $\widehat{N}_{R} = U^{\dag}_{R}N_{R}$,
so that the right-handed Majorana mass matrix $M_{R}$ becomes a diagonal matrix $\widehat{M}_R$ with real and positive mass eigenvalues $M_{1}=a M$, $M_{2}=M$ and $M_{3}=b M$,
 \begin{eqnarray}
  \widehat{M}_{R}&=& U^{T}_{R}M_{R}U_{R}=MU^{T}_{R}{\left(\begin{array}{ccc}
 1 &  0 &  0 \\
 0 &  1 &  \kappa e^{i\xi} \\
 0 &  \kappa e^{i\xi} &  1
 \end{array}\right)}U_{R}
= \begin{pmatrix} aM & 0 & 0 \\ 0 & M & 0 \\ 0 & 0 & bM \end{pmatrix},
  \label{heavy}
 \end{eqnarray}
where $\kappa=y^{\nu}_R v_{\chi}/M$. We find $a=\sqrt{1+\kappa^{2}+2\kappa\cos\xi}$, $b=\sqrt{1+\kappa^{2}-2\kappa\cos\xi}$, and a diagonalizing matrix
\begin{eqnarray}
  U_{R} = \frac{1}{\sqrt{2}}{\left(\begin{array}{ccc}
  0  &  \sqrt{2}  &  0 \\
  1 &  0  &  -1 \\
  1 &  0  &  1
  \end{array}\right)}{\left(\begin{array}{ccc}
  e^{i\frac{\psi_1}{2}}  &  0  &  0 \\
  0  &  1  &  0 \\
  0  &  0  &  e^{i\frac{\psi_2}{2}}
  \end{array}\right)}~,
  \label{URN}
\end{eqnarray}
with phases
\begin{eqnarray}
 \psi_1 = \tan^{-1} \Big( \frac{-\kappa\sin\xi}{1+\kappa\cos\xi} \Big)
 ~~~{\rm and}~~~ \psi_2 = \tan^{-1} \Big( \frac{\kappa\sin\xi}{1-\kappa\cos\xi} \Big)~.
\label{alphs_beta}
\end{eqnarray}
Interestingly, the mixing matrix of heavy neutrino $U_{R}$ in Eq.~(\ref{URN}) reflects an exact TBM.
As the magnitude of $\kappa$ defined in Eq.~(\ref{heavy}) decreases, the phases $\psi_{1,2}$ go to $0$ or $\pi$. And the Dirac neutrino mass term gets modified to $m_{D}\rightarrow \widetilde{m}_D = m_D U_R$ :
\begin{align}
\widetilde{m}_D=v_{\Phi}y^{\nu}_{1}\sqrt{\frac{3}{2}}{\left(\begin{array}{ccc}
 1 &  0 &  0 \\
 0 & y_{2} & 0 \\
 0 & 0 & y_{3}
 \end{array}\right)}{\left(\begin{array}{ccc}
 \sqrt{\frac{2}{3}} &  \frac{1}{\sqrt{3}} &  0 \\
 -\frac{1}{\sqrt{6}} & \frac{1}{\sqrt{3}} & -\frac{1}{\sqrt{2}} \\
 -\frac{1}{\sqrt{6}} & \frac{1}{\sqrt{3}} & -\frac{1}{\sqrt{2}}
 \end{array}\right)}{\left(\begin{array}{ccc}
  e^{i\frac{\psi_1}{2}}  &  0  &  0 \\
  0  &  1  &  0 \\
  0  &  0  &  e^{i\frac{\pi+\psi_2}{2}}
  \end{array}\right)}\equiv\frac{v_{\Phi}}{\sqrt{2}}\tilde{Y}_{\nu}~.
\label{mnuT}
\end{align}
At this point,
\begin{align}
-{\cal L}_{mW} &= \frac{1}{2} \begin{pmatrix} \overline{\nu_L} & \overline{\widehat{N}^{c}_R} \end{pmatrix} \begin{pmatrix} 0 & \widetilde{m}_D \\ \widetilde{m}_D^T & \widehat{M}_R \end{pmatrix} \begin{pmatrix} \nu^{c}_L \\ \widehat{N}_R \end{pmatrix} + \overline{\ell_{L}}\mathcal{M}_{\ell}\ell_{R}+\frac{g}{\sqrt{2}}W^{-}_{\mu}\overline{\ell_{L}}\gamma^{\mu}\nu_{L}+\text{h.c.}
 \label{lagrangianB}
\end{align}

Now we take the limit of large $M$ (seesaw mechanism)  and focus on the mass matrix of the light neutrinos $M_{\nu}$,
\begin{align}
-{\cal L}_{mW} &= \frac{1}{2} \overline{\nu_L} \mathcal{M}_{\nu} \nu^{c}_L  + \overline{\ell_{L}}\mathcal{M}_{\ell}\ell_{R}+\frac{g}{\sqrt{2}}W^{-}_{\mu}\overline{\ell_{L}}\gamma^{\mu}\nu_{L}+\text{h.c.}+\text{terms in $N_R$}
\end{align}
with
 \begin{align}
 \mathcal{M}_{\nu} = - \widetilde{m}_D \, \widehat{M}_R^{-1} \, \widetilde{m}^T_D.
 \end{align}
 We perform basis rotations from weak  to mass eigenstates in the leptonic sector,
\begin{eqnarray}
 \widehat{\ell}_{L} = P^{\ast}_{\ell}\ell_{L}~,\quad \widehat{\ell}_{R}= P^{\ast}_{\ell}\ell_{R}~,\quad \widehat{\nu}_{L} = U^{\dag}_{\nu}P^{\ast}_{\nu}\nu_{L}~,
 \label{rebasing}
\end{eqnarray}
where $P_{\ell}$ and $P_{\nu}$ are phase matrices and $U_\nu$ is a unitary matrix chosen so as the matrix
\begin{align}
\widehat{\mathcal{M}}_{\nu} = U^{\dag}_{\nu} P_\nu^* \mathcal{M}_\nu P_\nu^* U^*_{\nu} = - U_{\nu}^{\dag}P_{\nu}^{\ast} m_D U_R   \widehat{M}_R^{-1} (U_{\nu}^{\dag}P_{\nu}^{\ast} m_D U_R)^T
\end{align}
is diagonal.
Then from the charged current term in Eq.~(\ref{lagrangianB}) we obtain the lepton mixing matrix $U_{\rm PMNS}$ as
\begin{align}
U_{\rm PMNS}=P^{\ast}_{\ell}P_{\nu}U_{\nu}.
\end{align}
It is important to notice that the phase matrix $P_{\nu}$ can be rotated away by choosing the matrix $P_{\ell}=P_\nu$, i.e.\ by an appropriate redefinition of the left-handed charged lepton fields, which is always possible. This is an important point because the phase matrix $P_{\nu}$ accompanies the Dirac-neutrino mass matrix $\tilde{m}_D$ and ultimately the neutrino Yukawa matrix $Y_{\nu}$ in Eq.~(\ref{eq:Ynu}). This means that complex phases in $Y_{\nu}$ can always be rotated away by appropriately choosing the phases of left-handed charged lepton fields.
The matrix $U_{\rm PMNS}$ can be written in terms of three mixing angles and three $CP$-odd phases (one for the Dirac neutrinos and two for the Majorana neutrinos) as \cite{PDG}
\begin{eqnarray}
  U_{\rm PMNS}
  &=&{\left(\begin{array}{ccc}
   c_{13}c_{12} & c_{13}s_{12} & s_{13}e^{-i\delta_{CP}} \\
   -c_{23}s_{12}-s_{23}c_{12}s_{13}e^{i\delta_{CP}} & c_{23}c_{12}-s_{23}s_{12}s_{13}e^{i\delta_{CP}} & s_{23}c_{13}  \\
   s_{23}s_{12}-c_{23}c_{12}s_{13}e^{i\delta_{CP}} & -s_{23}c_{12}-c_{23}s_{12}s_{13}e^{i\delta_{CP}} & c_{23}c_{13}
   \end{array}\right)}Q_{\nu}~,
 \label{PMNS1}
\end{eqnarray}
where $Q_{\nu}={\rm Diag}(e^{-i\varphi_{1}/2},e^{-i\varphi_{2}/2},1)$, $s_{ij}\equiv \sin\theta_{ij}$ and $c_{ij}\equiv \cos\theta_{ij}$.

After seesawing, in a basis where charged lepton and heavy neutrino masses are real and diagonal, the light neutrino mass matrix is given by
 \begin{eqnarray}
  \mathcal{M}_{\nu} &=& -\widetilde{m}_{D}\widehat{M}^{-1}_{R}\widetilde{m}^{T}_{D} =-\frac{v^{2}_{\Phi}}{2}Y_{\nu}U_{R}\widehat{M}^{-1}_{R}U^{T}_{R}Y^{T}_{\nu}\nonumber\\
   &=& e^{i\pi}m_{0}
   {\left(\begin{array}{ccc}
   1+\frac{2e^{i\psi_{1}}}{a} & (1-\frac{e^{i\psi_{1}}}{a})y_{2} & (1-\frac{e^{i\psi_{1}}}{a})y_{3} \\
   (1-\frac{e^{i\psi_{1}}}{a})y_{2} & (1+\frac{e^{i\psi_{1}}}{2a}-\frac{3e^{i\psi_{2}}}{2b})y^{2}_{2} & (1+\frac{e^{i\psi_{1}}}{2a}+\frac{3e^{i\psi_{2}}}{2b})y_{2}y_{3}  \\
   (1-\frac{e^{i\psi_{1}}}{a})y_{3} & (1+\frac{e^{i\psi_{1}}}{2a}+\frac{3e^{i\psi_{2}}}{2b})y_{2}y_{3} & (1+\frac{e^{i\psi_{1}}}{2a}-\frac{3e^{i\psi_{2}}}{2b})y^2_{3}
   \end{array}\right)}~,
  \label{meff}
 \end{eqnarray}
where we have defined an overall scale $m_{0}=v^{2}_{\Phi}y^{\nu2}_{1}/(2M)$ for the light neutrino masses.
The mass matrix $\mathcal{M}_\nu$ is diagonalized by the PMNS mixing matrix $U_{\rm PMNS}$ as described above,
 \begin{eqnarray}
  \mathcal{M}_{\nu} &=& U_{\rm PMNS} ~{\rm Diag}(m_{1},m_{2},m_{3})~ U^{T}_{\rm PMNS} .
 \end{eqnarray}
Here $m_i$ $(i = 1,2,3)$ are the light neutrino masses. As is well known, because of the observed hierarchy $|\Delta m^{2}_{\rm Atm}|\equiv |m^{2}_{3}-m^{2}_{1}|\gg\Delta m^{2}_{\rm Sol}\equiv m^{2}_{2}-m^{2}_{1}>0$, and the requirement of a Mikheyev-Smirnov-Wolfenstein resonance for solar neutrinos, there are two possible neutrino mass spectra: (i) the normal mass hierarchy (NMH) $m_{1}<m_{2}<m_{3}$, and (ii) the inverted mass hierarchy (IMH) $m_{3}<m_{1}<m_{2}$.

In the limit $y^\nu_2=y^\nu_3$ ($y_{2}\rightarrow y_{3}$), the mass matrix in Eq.~(\ref{meff}) acquires a $\mu$--$\tau$ symmetry~\cite{mutau} that leads to $\theta_{13}=0$ and $\theta_{23}=-\pi/4$. Moreover, in the limit  $y^\nu_1=y^\nu_2=y^\nu_3$ ($y_{2}, y_{3}\rightarrow1$), the  mass matrix~(\ref{meff}) gives the TBM angles and their corresponding mass eigenvalues
 \begin{eqnarray}
\theta_{13}&=&0,\qquad\theta_{23}=\frac{\pi}{4}=45^\circ~,\qquad\theta_{12}=\sin^{-1}\left(\frac{1}{\sqrt{3}}\right)\simeq35.3^\circ~,\nonumber\\
 m_{1}&=& \frac{3m_{0}}{a}~,\qquad m_{2}=3m_{0}~,\qquad m_{3}= \frac{3m_{0}}{b}~.
 \label{TBM1}
 \end{eqnarray}
 These mass eigenvalues are disconnected from the mixing angles. However, recent neutrino data, {\it i.e.} $\theta_{13}\neq0$, require deviations of $y_{2,3}$ from unity, leading to a possibility to search for $CP$ violation in neutrino oscillation experiments. Eq.~(\ref{meff}) directly indicates that there could be deviations from the exact TBM if the Dirac neutrino Yukawa couplings do not have the same magnitude.
These deviations generate relations between mixing angles and mass eigenvalues.

To diagonalize the above mass matrix Eq.~(\ref{meff}), we consider the hermitian matrix $\mathcal{M}_{\nu}\mathcal{M}^{\dag}_{\nu}=U_{\nu}~{\rm Diag}(m^{2}_{1},m^{2}_{2},m^{2}_{3})~U^{\dag}_{\nu}$, from which we obtain the masses and mixing angles:
 \begin{eqnarray}
 \mathcal{M}_{\nu}\mathcal{M}^{\dag}_{\nu}=m^{4}_{0}\left(\begin{array}{ccc}
  A & y_{2}B & y_{3}C \\
  y_{2}B^{\ast} & y^{2}_{2}F & y_{2}y_{3}|G|e^{i\phi^\nu_{1}} \\
  y_{3}C^{\ast} & y_{2}y_{3}|G|e^{-i\phi^\nu_{1}} & y^{2}_{3}K
  \end{array}\right)=U_{\nu}~{\rm Diag}(m^{2}_{1},m^{2}_{2},m^{2}_{3})~U^{\dag}_{\nu}~,
 \label{MM}
 \end{eqnarray}
where the parameters $A,B,C,F,G$ and $K$ are given in Eq.~(\ref{mnu_elements}).
The mixing matrix $U_{\nu}$ in Eq.~(\ref{MM}) associated with diagonalization giving definite masses can be written as
 \begin{eqnarray}
 U_{\nu}=e^{i\Psi_{\nu}}{\left(\begin{array}{ccc}
 c_{2}c_{3} &  c_{2}s_{3}e^{i\phi^\nu_{3}} &  s_{2}e^{i\phi^\nu_{2}} \\
 -c_{1}s_{3}e^{-i\phi^\nu_{3}}-s_{1}s_{2}c_{3}e^{i(\phi^\nu_{1}-\phi^\nu_{2})} &  c_{1}c_{3}-s_{1}s_{2}s_{3}e^{i(\phi^\nu_{1}-\phi^\nu_{2}+\phi^\nu_{3})} &  s_{1}c_{2}e^{i\phi^\nu_{1}} \\
 s_{1}s_{3}e^{-i(\phi^\nu_{1}+\phi^\nu_{3})}-c_{1}s_{2}c_{3}e^{-i\phi^\nu_{2}} &  -s_{1}c_{3}e^{-i\phi^\nu_{1}}-c_{1}s_{2}s_{3}e^{i(\phi^\nu_{3}-\phi^\nu_{2})} &  c_{1}c_{2}
 \end{array}\right)}Q'_{\nu}~
 \label{Ugeneral}
 \end{eqnarray}
where $c_{i}\equiv\cos\theta_{i}$, $s_{i}\equiv\sin\theta_{i}$ and a diagonal phase matrix $Q'_{\nu}={\rm diag}(1,e^{i\zeta_{1}},e^{i\zeta_{2}})$.
Now, the straightforward calculation with the general
parametrization of $U_{\nu}$ in Eq.~(\ref{Ugeneral}) leads to the expressions for the masses and mixing angles~\cite{Ahn:2013ema}:
 \begin{eqnarray}
  \tan\theta_{1}&=&\frac{y_{3}}{y_{2}}\frac{{\rm Im}[C]\sin\phi^\nu_{2}-{\rm Re}[C]\cos\phi^\nu_{2}}{{\rm Im}[B]\cos(\phi^\nu_{1}-\phi^\nu_{2})+{\rm Re}[B]\sin(\phi^\nu_{1}-\phi^\nu_{2})}~, \qquad \phi^\nu_{1}=\arg(G)~,\nonumber\\ \tan2\theta_{2}&=&2\frac{|c_{1}y_{3}C+e^{i\phi^\nu_{1}}s_{1}y_{2}B|}{\lambda_{3}-A}~,\qquad\phi^\nu_{2}=\arg\left(c_{1}y_{3}C+e^{i\phi^\nu_{1}}s_{1}y_{2}B\right)~,\nonumber\\
  \tan2\theta_{3}&=&2\frac{|Z|}{\lambda_{2}-\lambda_{1}}~,~\quad \phi^\nu_{3}=\arg(Z)~,
 \label{Theta1312}
 \end{eqnarray}
where
 \begin{eqnarray}
  \lambda_{1}&=&Ac^{2}_{2}-|c_{1}y_{3}C+e^{i\phi^\nu_{1}}s_{1}y_{2}B|\sin2\theta_{2}+\lambda_{3}s^{2}_{2}~,\nonumber\\
  \lambda_{2}&=&y^{2}_{2}Fc^{2}_{1}-y_{2}y_{3}|\tilde{G}|\sin2\theta_{1}+y^{2}_{3}Ks^{2}_{1}~,\qquad
  \lambda_{3}=y^{2}_{3}Kc^{2}_{1}+y_{2}y_{3}|\tilde{G}|\sin2\theta_{1}+y^{2}_{2}Fs^{2}_{1}~,\nonumber\\
  Z&=& c_{2}(c_{1}y_{2}B-e^{-i\phi^\nu_{1}}s_{1}y_{3}C)+s_{2}e^{i(\phi^\nu_{2}-\phi^\nu_{1})}\left(\sin2\theta_{1}\frac{y^{2}_{3}K-y^{2}_{2}F}{2}-y_{2}y_{3}|G|\cos2\theta_{1}\right)~.
 \label{para1}
 \end{eqnarray}
And the squared-mass eigenvalues are given by
 \begin{eqnarray}
    m^{2}_{1} &=& m^{2}_{0}\frac{\lambda_{1}c^{2}_{3}-\lambda_{2}s^{2}_{3}}{\cos2\theta_{3}}~,\qquad\qquad
    m^{2}_{2} = m^{2}_{0}\frac{\lambda_{2}c^{2}_{3}-\lambda_{1}s^{2}_{3}}{\cos2\theta_{3}}~,\nonumber\\
    m^{2}_{3}&=&m^{2}_{0}\left(\lambda_{3}+|c_{1}y_{3}C+e^{i\phi^\nu_{1}}s_{1}y_{2}B|\tan\theta_{2}\right)~.
 \label{eigenvalueNu}
 \end{eqnarray}
Without loss of generality, we let $\theta_{1}\equiv\theta_{23}$, $\theta_{2}\equiv\theta_{13}$ and $\theta_{3}\equiv\theta_{12}$.
In the limit of $y_{2},y_{3}\rightarrow1$, the parameters relevant for mixing angles behave as $A,\lambda_{1}\rightarrow3\left(1+\frac{2}{a^2}\right)$, $\lambda_{2}\rightarrow6\left(1+\frac{1}{2a^2}\right)$, $B,C\rightarrow3\left(1-\frac{1}{a^2}\right)$, $Z\rightarrow\frac{6}{\sqrt{2}}\left(1-\frac{1}{a^2}\right)$, and which in turn imply $\theta_{23}\rightarrow-\pi/4$, $\phi^\nu_{1,2,3}\rightarrow0$, $\theta_{13}\rightarrow0$ and $\tan2\theta_{12}\rightarrow2\sqrt{2}$. So, the lifts of $y_{2}, y_{3}$ from unit or inequality between them can trigger deviations from the TBM.

Leptonic CP violation can be detected through the neutrino oscillations which are sensitive to the Dirac CP phase $\delta_{CP}$, but insensitive to the Majorana phases in $U_{\rm PMNS}$~\cite{Branco:2002xf}.
To see how the parameters are correlated with low-energy CP violation observables measurable through neutrino oscillations, we consider the leptonic CP violation parameter defined by the
Jarlskog invariant~\cite{Jarlskog:1985ht} in the standard parametrization~Eq.~(\ref{PMNS1}):
 \begin{eqnarray}
 J_{CP}\equiv-{\rm Im}[U^{\ast}_{e1}U_{e3}U_{\tau1}U^{\ast}_{\tau3}]=\frac{1}{8}\sin2\theta_{12}\sin2\theta_{13}\sin2\theta_{23}\cos\theta_{13}\sin\delta_{CP}~,
 \label{rephasing}
 \end{eqnarray}
where $U_{\alpha j}$ is an element of the PMNS matrix in Eq.~(\ref{PMNS1}), with $\alpha=e,\mu,\tau$
corresponding to the lepton flavors and $j=1,2,3$ corresponding to the light neutrino mass eigenstates.
At the same time, in the parametrization given in Eq.~(\ref{Ugeneral}) we obtain
 \begin{eqnarray}
 J_{CP}=\frac{1}{8}\sin2\theta_{1}\sin2\theta_{2}\sin2\theta_{3}\cos\theta_{2}\sin(\phi^\nu_{1}-\phi^\nu_{2}+\phi^\nu_{3})~.
 \label{rephasing1}
 \end{eqnarray}
From Eqs.~(\ref{rephasing}) and (\ref{rephasing1}) we obtain the Dirac CP phase defined in Eq.~(\ref{PMNS1}) as
 \begin{eqnarray}
 \delta_{CP}=\phi^\nu_{1}-\phi^\nu_{2}+\phi^\nu_{3}~.
 \label{DiracCP}
 \end{eqnarray}
The phase $\phi^\nu_{i}$ ($i=1,2,3$) or $\delta_{CP}$ is constrained by the neutrino mass matrix Eq.~(\ref{meff}), which is originated from the phase $\xi$.
The Jarlskog invariant $J_{CP}$ can be expressed in terms of the elements of the matrix $h=\mathcal{M}_{\nu}\mathcal{M}^{\dag}_{\nu}$~\cite{Branco:2002xf}:
 \begin{eqnarray}
  J_{CP}=-\frac{{\rm Im}\{h_{12}h_{23}h_{31}\}}{\Delta m^{2}_{21}\Delta m^{2}_{31}\Delta m^{2}_{32}}~,
  \label{JCP}
 \end{eqnarray}
where the numerator is expressed as
 \begin{eqnarray}
  {\rm Im}\{h_{12}h_{23}h_{31}\}&=&m^{6}_{0}\frac{27y^{2}_{2}y^{2}_{3}(y^{2}_{2}-y^{2}_{3})}{2}\Big(\sin(\psi_{1}-\psi_{2})\{....\}+\sin(2\psi_{1}-\psi_{2})\{.....\}\nonumber\\
  &+&\sin\psi_{2}\{....\}+\sin(\psi_{1}+\psi_{2})\{....\}\Big)~,
  \label{JCP1}
 \end{eqnarray}
in which $\{.....\}$ stands for a complicated lengthy function of $y_{2}$, $y_{3}$, $a$ and $b$. Clearly, Eq.~(\ref{JCP1}) indicates that $J_{CP}$ depends on the phase $\xi$ (or $\psi_{1,2}$) and, in the limit of $y_{2}\rightarrow y_{3}$, the leptonic CP violation $J_{CP}$ goes to zero.

Concerning CP violation, we notice that the CP phase $\xi$ coming from $M_{R}$ take part in low-energy CP violation in terms of $\psi_1,\psi_2$, as can be seen in Eqs.~(\ref{heavy}-\ref{meff}). Any CP-violation relevant for leptogenesis is associated with the neutrino Yukawa matrix $\widetilde{Y}_{\nu}=Y_{\nu}U_{R}$ and the combination of Dirac neutrino Yukawa matrices, $H\equiv\widetilde{Y}^{\dag}_{\nu}\widetilde{Y}_{\nu}= U^{\dag}_{R}Y^{\dag}_{\nu}Y_{\nu}U_{R}$, which is
 \begin{eqnarray}
   H=|y^{\nu}_{1}|^{2}\left(\begin{array}{ccc}
  \frac{4+y^{2}_{2}+y^{2}_{3}}{2} & \frac{e^{-i\frac{\psi_{1}}{2}}}{\sqrt{2}}(2-y^{2}_{2}-y^{2}_{3}) & -\frac{i\sqrt{3}e^{i\frac{\psi_{21}}{2}}}{2}(y^{2}_{2}-y^{2}_{3}) \\
  \frac{e^{i\frac{\psi_{1}}{2}}}{\sqrt{2}}(2-y^{2}_{2}-y^{2}_{3}) & 1+y^{2}_{2}+y^{2}_{3} & i\sqrt{\frac{3}{2}}e^{i\frac{\psi_{2}}{2}}(y^{2}_{2}-y^{2}_{3}) \\
  \frac{i\sqrt{3}e^{-i\frac{\psi_{21}}{2}}}{2}(y^{2}_{2}-y^{2}_{3}) & -i\sqrt{\frac{3}{2}}e^{-i\frac{\psi_{2}}{2}}(y^{2}_{2}-y^{2}_{3}) & \frac{3}{2}(y^{2}_{2}+y^{2}_{3})
  \end{array}\right) ,
 \label{YnuYnu}
 \end{eqnarray}
where $\psi_{ij}\equiv\psi_{i}-\psi_{j}$. As expected, in the limit $y^{\nu}_1=y^{\nu}_2=y^{\nu}_3$ , i.e.\ $y_{2,3}\rightarrow1$, the off-diagonal entries of $H$  vanish, and there is no CP violation useful for leptogenesis. If the Dirac neutrino Yukawa couplings $y^{\nu}_1$, $y^{\nu}_2$, and $y^{\nu}_3$ differ in magnitude, they can play a role in baryogenesis via leptogenesis and nonzero $\theta_{13}\simeq9^{\circ}$ with two large mixing angles ($\theta_{23},\theta_{12}$). Therefore,
a low energy CP violation in neutrino oscillation and/or a high energy CP
violation in leptogenesis can be generated by the non-degeneracy of the Dirac neutrino Yukawa couplings and a nonzero phase $\xi$ coming from $M_{R}$.

In summary, the phase $\xi$ originated from the heavy gauge singlet $\chi$ field is responsible for leptogenesis, a CP phase in neutrino oscillation, $\delta_{CP}$, and the Dirac CP phase in the CKM mixing matrix, $\delta^{q}_{CP}$.

\section{Numerical Study}
Now we perform a numerical analysis using the linear algebra tools in Ref.~\cite{Antusch:2005gp}.
The Daya Bay and RENO experiments have
accomplished the measurement of three mixing angles
$\theta_{12}, \theta_{23}$, and $\theta_{13}$ from three kinds of neutrino oscillation experiments.
The global fit of the neutrino mixing angles and of the mass-squared differences at the $1\sigma$ $(3\sigma)$ level is
given by~\cite{GonzalezGarcia:2012sz}
 \begin{eqnarray}
  &&\theta_{13}=8.66^{\circ+0.44^{\circ}~(+1.30^{\circ})}_{~-0.46^{\circ}~(-1.47^{\circ})},\qquad\delta_{\rm CP}=300^{\circ+66^{\circ}~~(+60^{\circ})}_{~-138^{\circ}~(-300^{\circ})},\qquad\theta_{12}=33.36^{\circ+0.81^{\circ}~(+2.53^{\circ})}_{~-0.78^{\circ}~(-1.27^{\circ})},\nonumber\\
  &&\theta_{23}=40.0^{\circ+2.1^{\circ}}_{~-1.5^{\circ}}\oplus50.4^{\circ+1.3^{\circ}}_{~-1.3^{\circ}}~{1\sigma},\quad\left(35.8^{\circ}\thicksim54.8^{\circ}~{3\sigma}\right),\nonumber
 \end{eqnarray}
 \begin{eqnarray}
  &&\Delta m^{2}_{\rm Sol}[10^{-5}{\rm eV}^{2}]=7.50^{+0.18~(+0.59)}_{-0.19~(-0.50)},~\Delta m^{2}_{\rm Atm}[10^{-3}{\rm eV}^{2}]=\left\{\begin{array}{ll}
                2.473^{+0.070~(+0.222)}_{-0.067~(-0.197)}, & \hbox{NMH} \\
                2.427^{+0.042~(+0.185)}_{-0.065~(-0.222)}, & \hbox{IMH}~
                                  \end{array},
                                \right.
 \label{expnu}
 \end{eqnarray}
where $\Delta m^{2}_{\rm Sol}\equiv m^{2}_{2}-m^{2}_{1}$, $\Delta m^{2}_{\rm Atm}\equiv m^{2}_{3}-m^{2}_{1}$ for the normal mass hierarchy (NMH), and  $\Delta m^{2}_{\rm Atm}\equiv |m^{2}_{3}-m^{2}_{2}|$ for the inverted mass hierarchy (IMH).
The matrices $m_{D}$ and $\hat{M}_{R}$ in Eq.~(\ref{meff}) contain seven parameters : $y^{\nu}_{1},M,v_{\Phi},y_{2},y_{3},\kappa,\xi$. The first three ($y^{\nu}_{1}$, $M,$ and $v_{\Phi}$) lead to the overall neutrino scale parameter $m_{0}$. The next four ($y_2,y_3,\kappa,\xi$) give rise to the deviations from TBM as well as the CP phases and corrections to the mass eigenvalues (see Eq.~(\ref{TBM1})).

In our numerical examples, we take $M=10^{11}$ GeV and $v_{\eta}=v_{\Phi}=123$ GeV, for simplicity, as inputs~\footnote{If one takes a seesaw scale $M=10^{11}$ GeV, then the cutoff scale would be around $10^{12}$ GeV due to the relation $v_{\chi}/\Lambda=\lambda$ in Eqs.~(\ref{vevhier}), (\ref{DL}) and (\ref{ckm1}).}. Since the neutrino masses are sensitive to the combination $m_{0}=v^{2}_{\Phi}|y^{\nu}_{1}|^{2}/(2M)$, other choices of $M$ and $v_\Phi$ give identical results. Then the parameters $m_{0},y_{2},y_{3},\kappa,\xi$ can be determined from the experimental results of three mixing angles, $\theta_{12},\theta_{13},\theta_{23}$, and the
two mass squared differences, $\Delta m^{2}_{\rm Sol}, \Delta m^{2}_{\rm Atm}$. In addition, the CP phases $\delta_{CP},\varphi_{1,2}$ can be predicted after determining the model parameters.

Using the formulas for the neutrino mixing angles and masses and our values of $M,v_\eta,v_\Phi$, we obtain the following allowed regions of the unknown model parameters: for the normal mass hierarchy (NMH)~\footnote{When $y_{2}=y_{3}$ and around there, there exist other parameter spaces giving very small values of $\theta_{13}$. So, we have neglected them in our numerical result for normal mass hierarchy.},
 \begin{eqnarray}
  &&0.17\lesssim\kappa\lesssim0.80,\qquad1.0\lesssim y_{2}\lesssim1.3,\qquad1.0\lesssim y_{3}<1.3,\nonumber\\
  &&99^{\circ}\lesssim\xi\lesssim117^{\circ},\qquad
        243^{\circ}\lesssim\xi\lesssim263^{\circ}
  ,\qquad
1.3\lesssim m_{0}\times10^{-2}{\rm [eV]}\lesssim4.3;
  \label{input1}
 \end{eqnarray}
 for the inverted mass hierarchy (IMH),
 \begin{eqnarray}
  &&0.3\lesssim\kappa\lesssim1.9,\qquad0.70\lesssim y_{2}\lesssim1.30,\qquad0.74\lesssim y_{3}\lesssim1.31,\nonumber\\
  && 96^{\circ}\lesssim\xi\lesssim160^{\circ},\qquad
        212^{\circ}\lesssim\xi\lesssim265^{\circ}
  ,\qquad1.6\lesssim m_{0}\times10^{-2}{\rm [eV]}\lesssim4.0.
  \label{input2}
 \end{eqnarray}
Note that here we have used the $3\sigma$ experimental bounds on $\theta_{12},\theta_{23},\Delta m^{2}_{\rm Sol}, \Delta m^{2}_{\rm Atm}$ in Eq.~(\ref{expnu}), except for $\theta_{13}<11^{\circ}$ for which we use the values in Eqs.~(\ref{input1},\ref{input2}).

\begin{figure}[t]
\begin{minipage}[t]{6.0cm}
\epsfig{figure=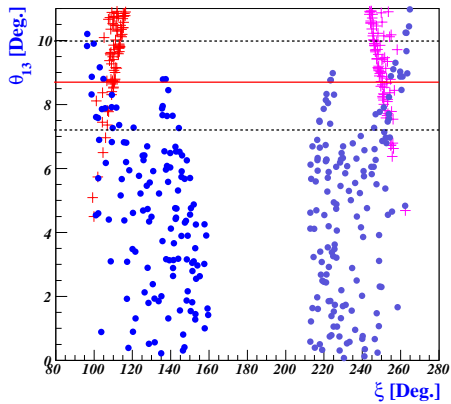,width=6.5cm,angle=0}
\end{minipage}
\caption{\label{FigA1pp}
The reactor mixing angle $\theta_{13}$ versus the phase $\xi$ of the parameter combination $y^{\nu}_R v_\chi/M$. The horizontal dotted (solid) lines in both plots indicate the upper and lower bounds at the $3\sigma$ level (the best-fit value) on $\theta_{13}$ given in Eq.~(\ref{expnu}). The red-type crosses and blue-type dots represent the results for the normal and inverted mass hierarchy, respectively; the data points of red-crosses and blue-dots corresponding to $95^{\circ}\lesssim\xi\lesssim145^{\circ}$ within 3$\sigma$ experimental bounds of $\theta_{13}$ can explain the CKM CP phase.}
\end{figure}
\begin{figure}[t]
\begin{minipage}[t]{6.0cm}
\epsfig{figure=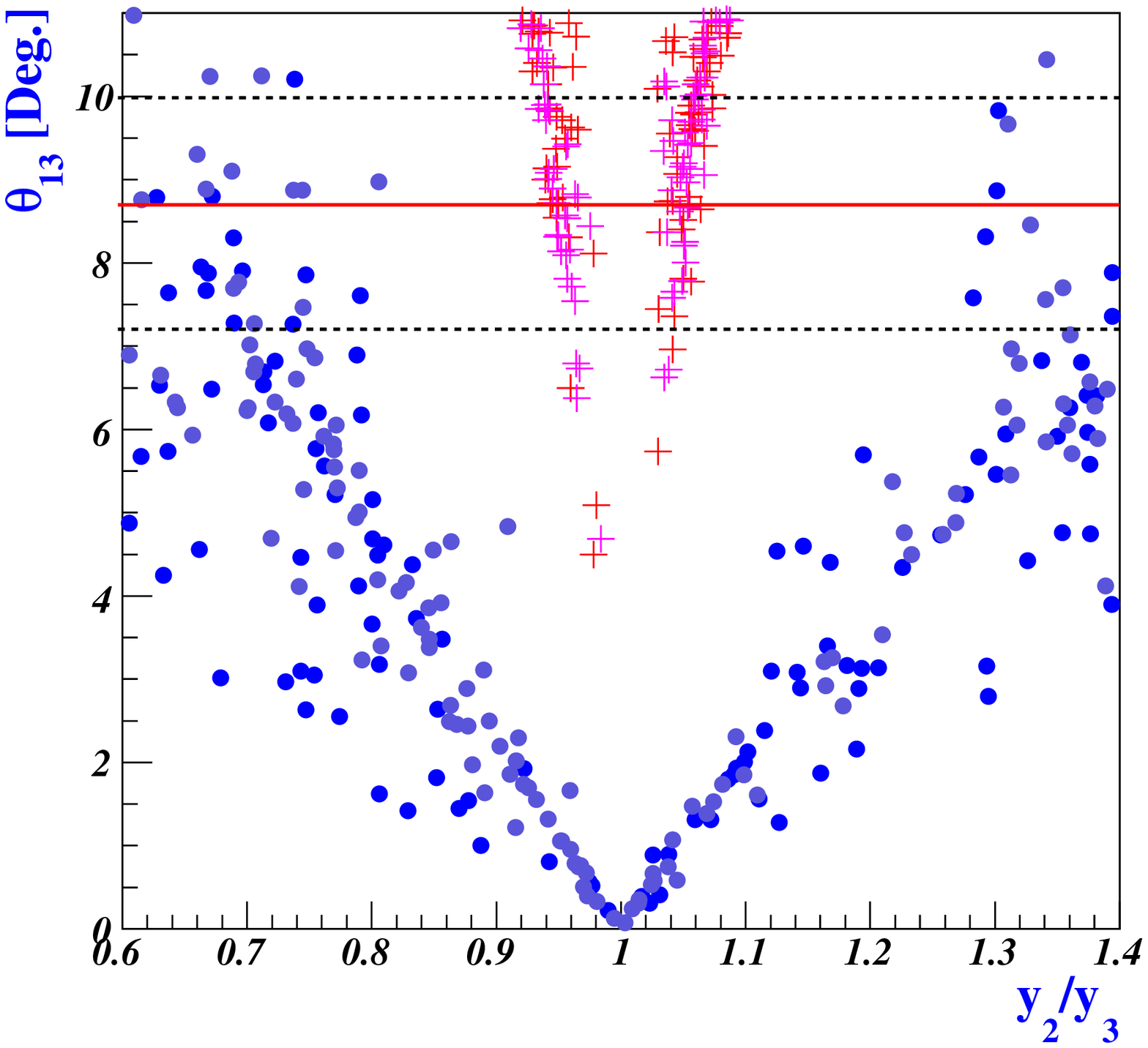,width=6.5cm,angle=0}
\end{minipage}
\hspace*{1.0cm}
\begin{minipage}[t]{6.0cm}
\epsfig{figure=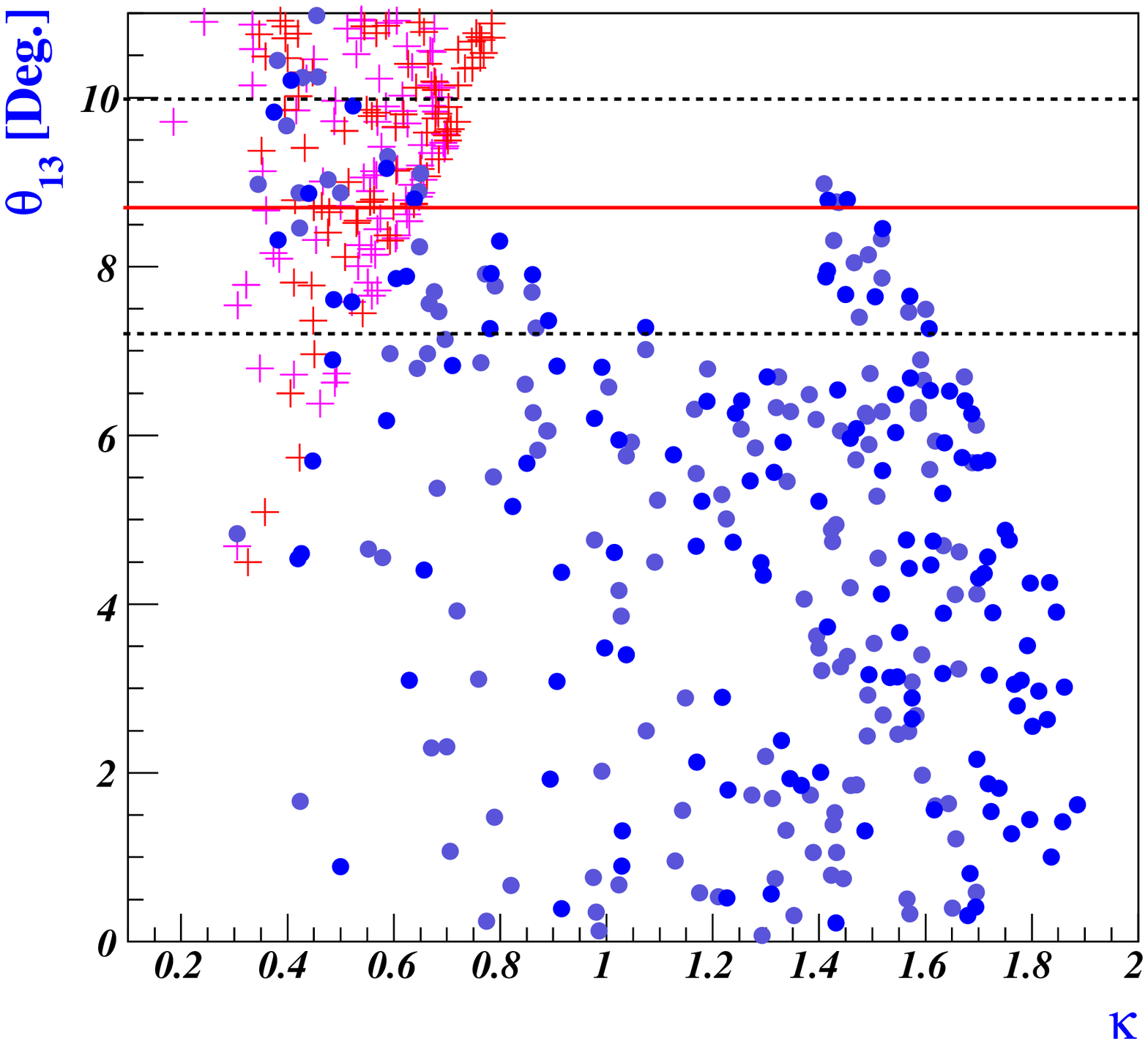,width=6.5cm,angle=0}
\end{minipage}
\caption{\label{FigA1}
The reactor mixing angle $\theta_{13}$ versus the ratio of second-to-third generation neutrino Yukawa couplings $y^{\nu}_{2}/y^{\nu}_{3}$ (left plot) and the parameter $\kappa=y^{\nu}_{R} v_\chi/M$ (right plot). The red-type crosses and blue-type dots represent the results for the normal and the inverted mass hierarchy, respectively. The horizontal dotted (solid) lines in both plots indicate the upper and lower bounds at the $3\sigma$ level (the best-fit value) on $\theta_{13}$ given in Eq.~(\ref{expnu}).}
\end{figure}
\begin{figure}[t]
\begin{minipage}[t]{6.0cm}
\epsfig{figure=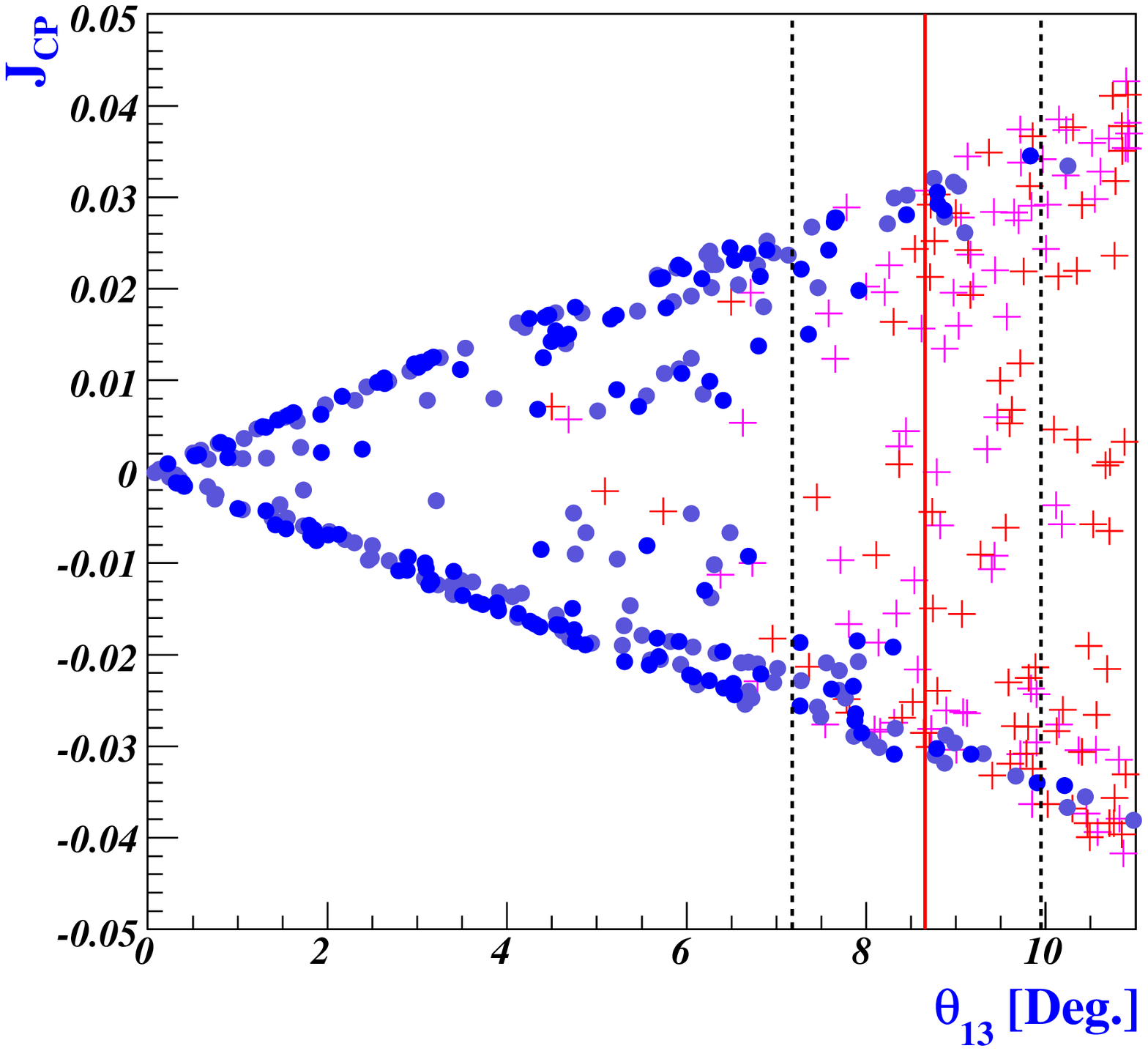,width=6.5cm,angle=0}
\end{minipage}
\hspace*{1.0cm}
\begin{minipage}[t]{6.0cm}
\epsfig{figure=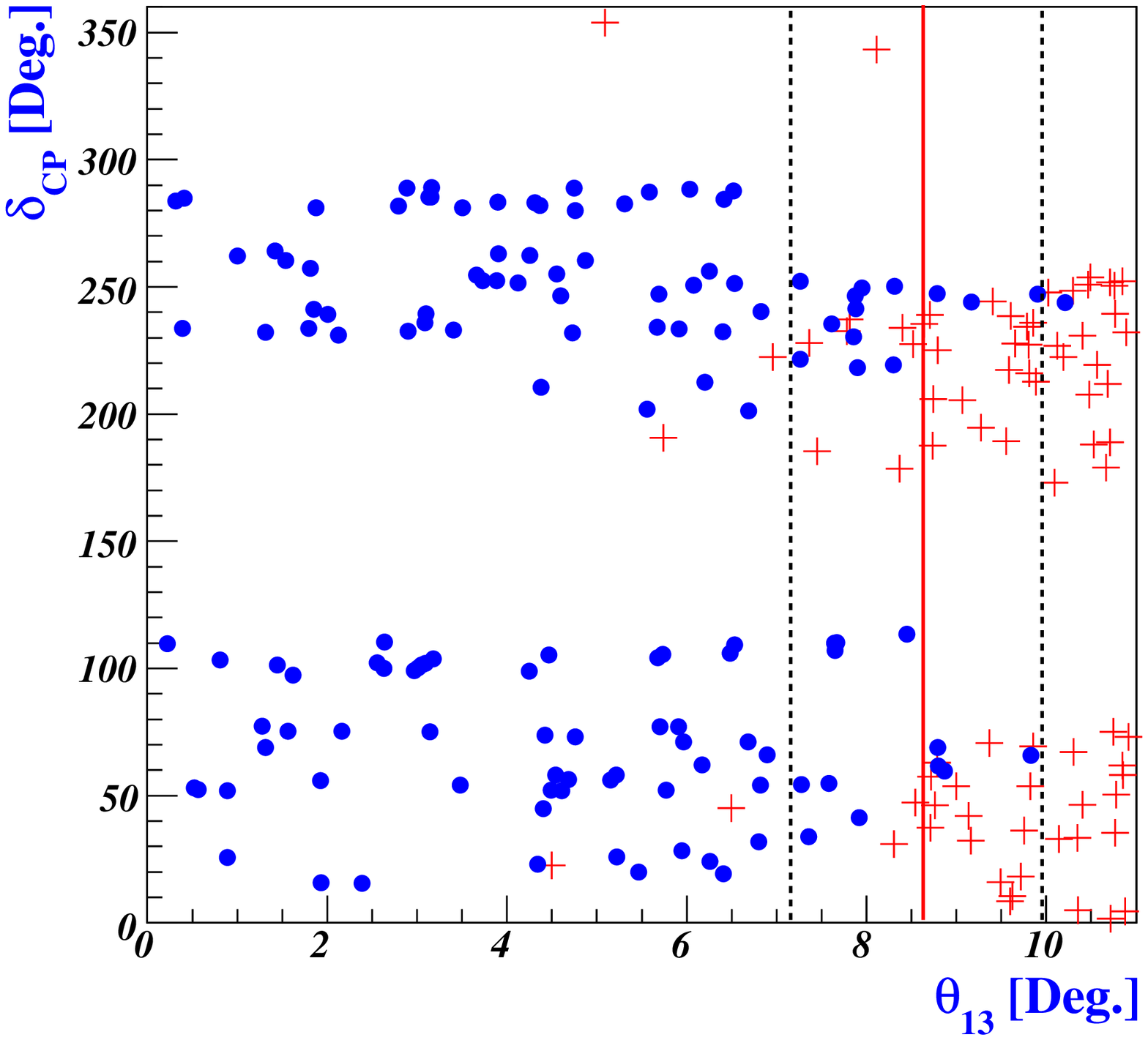,width=6.5cm,angle=0}
\end{minipage}
\caption{\label{FigA1p}
The Jarlskog invariant $J_{CP}$ versus the reactor angle $\theta_{13}$ (left plot), and the Dirac CP phase $\delta_{CP}$ versus $\theta_{13}$ satisfying the measured CKM phase (right plot). The vertical dotted (solid) lines in both plots indicate the upper and lower bounds at the $3\sigma$ level (the best-fit value)  on $\theta_{13}$ given in Eq.~(\ref{expnu}).}
\end{figure}
\begin{figure}[t]
\begin{minipage}[t]{6.0cm}
\epsfig{figure=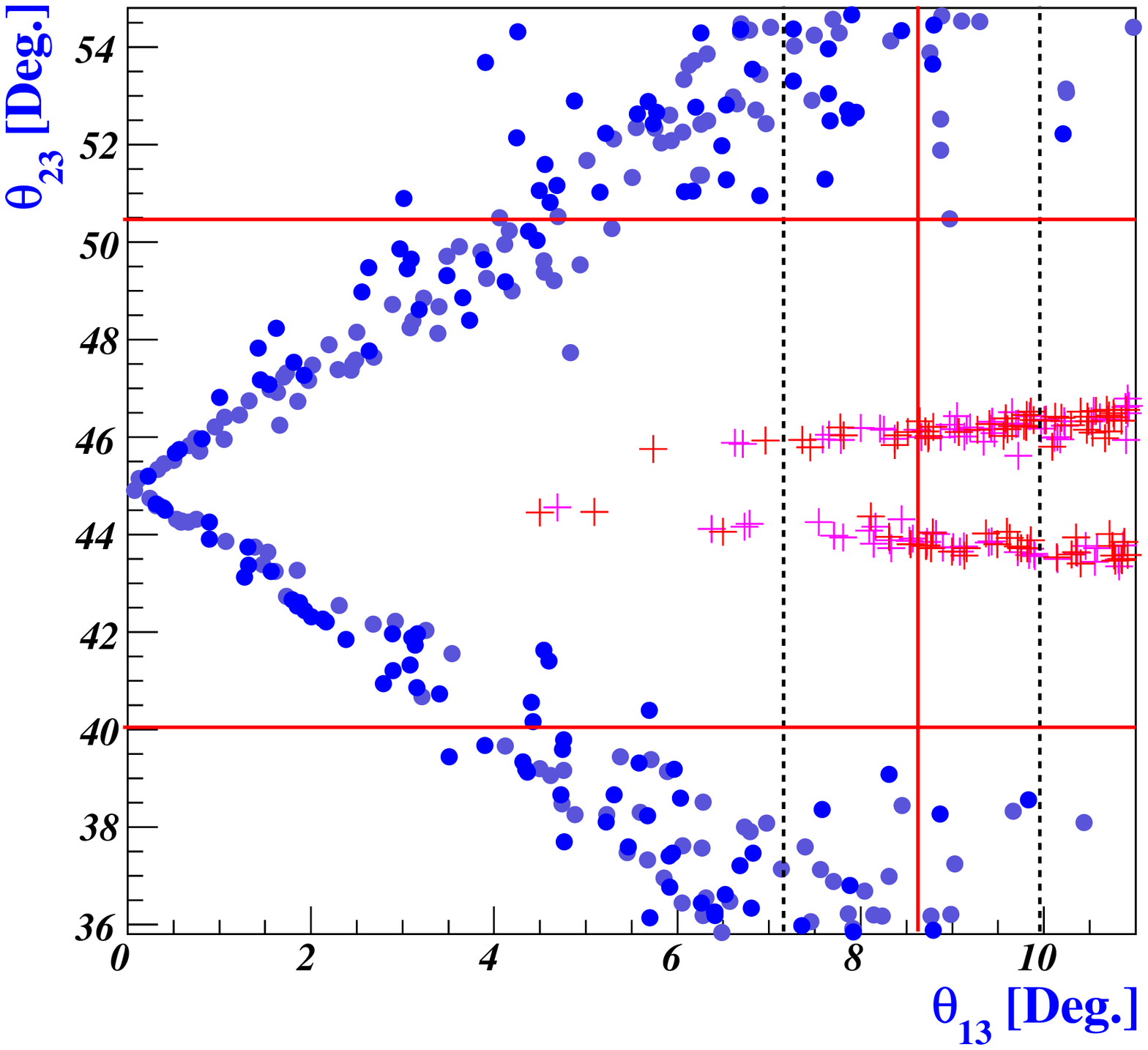,width=6.5cm,angle=0}
\end{minipage}
\hspace*{1.0cm}
\begin{minipage}[t]{6.0cm}
\epsfig{figure=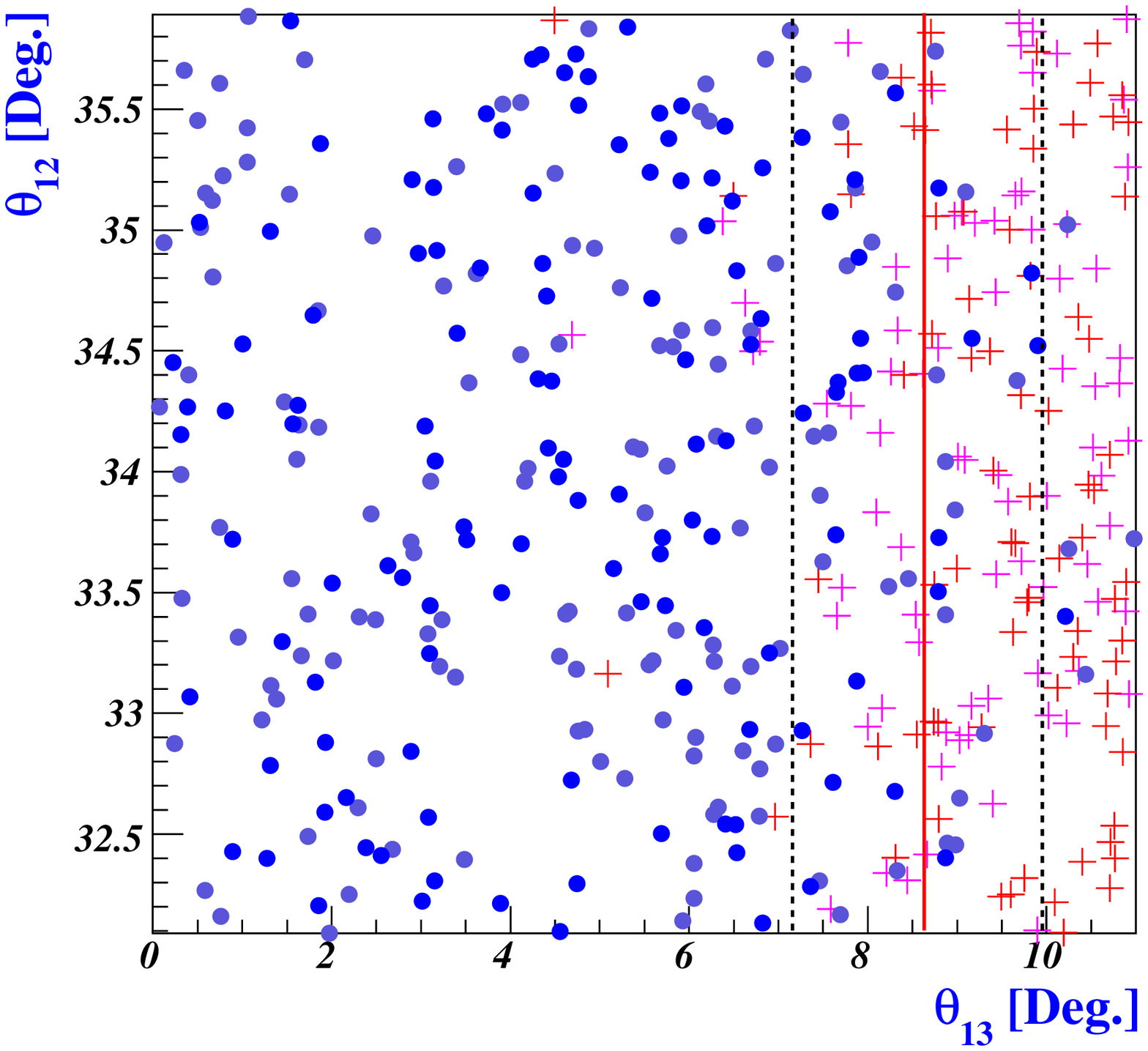,width=6.5cm,angle=0}
\end{minipage}
\caption{\label{FigA2} The behaviors of $\theta_{23}$ and $\theta_{12}$ in terms of  $\theta_{13}$. The dotted vertical lines represent the experimental bounds of Eq.~(\ref{expnu}) at $3\sigma$'s. The horizontal and vertical solid lines indicate the $1\sigma$ experimental best-fit values in Eq.~(\ref{expnu}).
}
\end{figure}
\begin{figure}[t]
\begin{minipage}[t]{6.0cm}
\epsfig{figure=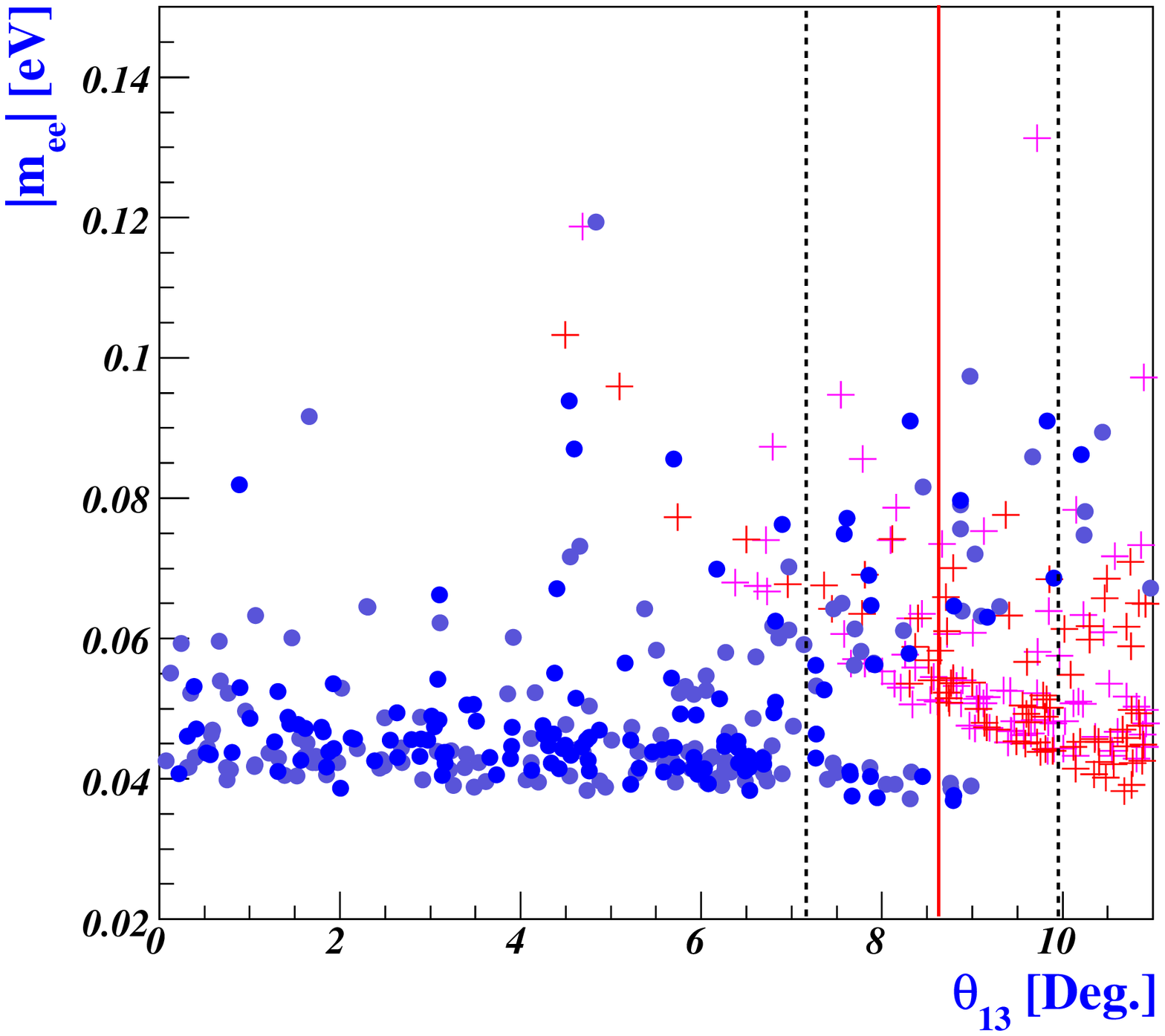,width=6.5cm,angle=0}
\end{minipage}
\hspace*{1.0cm}
\begin{minipage}[t]{6.0cm}
\epsfig{figure=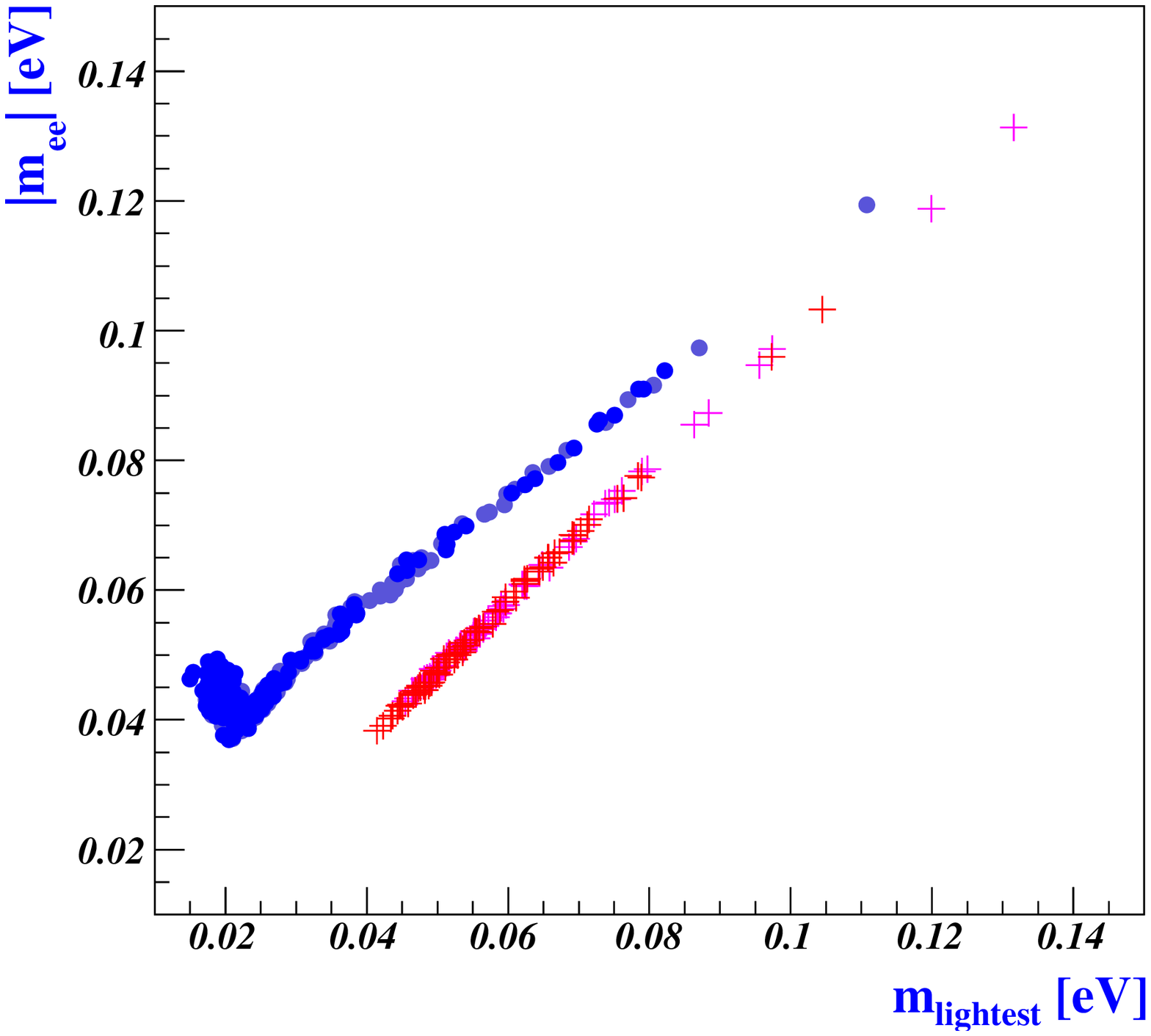,width=6.5cm,angle=0}
\end{minipage}
\caption{\label{FigA3} Plots of $|m_{ee}|$ as a function of $\theta_{13}$ and $m_{\rm lightest}$. The vertical dotted (solid) lines show the experimental bounds of Eq.~(\ref{expnu}) at $3\sigma$'s (the best-fit value).}
\end{figure}

For these parameter regions, we investigate how a nonzero $\theta_{13}$ can be determined for the normal and inverted mass hierarchy. In Figs.~\ref{FigA1pp}-\ref{FigA3}, the data points represented by blue-type dots and red-type crosses indicate results for the inverted and normal mass hierarchy, respectively. Fig.~\ref{FigA1pp} shows the reactor mixing angle $\theta_{13}$ as a function of the phase $\xi$. As can be seen in Fig.~\ref{FigA1pp}, the data points in ranges of $100^{\circ}\lesssim\xi\lesssim115^{\circ}$ (NMH) and $95^{\circ}\lesssim\xi\lesssim145^{\circ}$ (IMH) within $3\sigma$ experimental bounds of $\theta_{13}$ can explain the CKM CP phase as explained in Eqs.~(\ref{lagrangianA})-(\ref{CKMpara2}).
The left-hand-side plot in Fig.~\ref{FigA1} shows how the mixing angle $\theta_{13}$ depends on the ratio $y_2/y_3=y_2^{\nu}/y_3^{\nu}$ of the second- and third-generation neutrino Yukawa couplings; the right-hand-side plot shows how $\theta_{13}$ depends on the parameter $\kappa=y^{\nu}_R v_\chi/M$. We see that the measured value of $\theta_{13}$ from the Daya Bay and RENO experiments can be achieved at $3\sigma$'s for $0.93\lesssim y_{2}/y_{3}\lesssim0.98$ and $1.03\lesssim y_{2}/y_{3}\lesssim1.08$ (NMH), $0.6\lesssim y_{2}/y_{3}\lesssim0.82$ and $1.2\lesssim y_{2}/y_{3}\lesssim1.4$ (IMH), $0.17\lesssim\kappa\lesssim0.75$ (NMH) and $0.3<\kappa\lesssim1.1$ and $1.4\lesssim\kappa\lesssim1.6$ (IMH).

The behavior of $J_{CP}$ defined in Eqs.~(\ref{rephasing})-(\ref{JCP1}) as a function of $\theta_{13}$ is plotted on the left plot of Fig.~\ref{FigA1p}.
We see that the value of $|J_{CP}|$ lies in the range $0-0.038$ (NMH) and $0.014-0.034$ (IMH) for the measured value of $\theta_{13}$ at $3\sigma$'s.
When $y_{2}\neq1$, i.e.\ for the normal hierarchy case, $J_{CP}$ could go to zero as $\sin\psi_{2}$ of Eq.~(\ref{JCP1}).
In the case of the inverted hierarchy, $J_{CP}$ has nonzero values for the measured range of $\theta_{13}$ while $J_{CP}$ goes to zero for $\theta_{13}\rightarrow0$, which corresponds to $y_{2}\rightarrow1$.
Interestingly enough, the right plot of Fig.~\ref{FigA1p} shows that the data points satisfying the CKM CP phase favor the values around $60^{\circ},110^{\circ}$ and $230^{\circ}$ for the inverted mass hierarchy, and around $30^{\circ}$ and $200^{\circ}$ for normal mass one.

Fig.~\ref{FigA2} shows how the values of $\theta_{13}$ depend on the mixing angles $\theta_{23}$ and $\theta_{12}$. As can be seen in the left plot of Fig.~\ref{FigA2}, the behavior of $\theta_{23}$ in terms of the measured values of $\theta_{13}$ at $3\sigma$'s for the normal hierarchy is different than for the inverted hierarchy.  For the normal hierarchy we see that the measured values of $\theta_{13}$ can be achieved for $43.5^{\circ}\lesssim\theta_{23}\lesssim44.5^{\circ}$ and $45.5^{\circ}\lesssim\theta_{23}\lesssim47.0^{\circ}$ with small deviations from maximality, which are disfavored at $1\sigma$ by the experimental bounds  as can be seen in Eq.~(\ref{expnu}), while for the inverted hierarchy $50^{\circ}\lesssim\theta_{23}\lesssim54.8^{\circ}$ and $35.8^{\circ}\lesssim\theta_{23}\lesssim39^{\circ}$, which are favored at $1\sigma$ by the experimental bounds in Eq.~(\ref{expnu}).
From the right plot of Fig.~\ref{FigA2}, we see that the
predictions for $\theta_{13}$ do not strongly depend on $\theta_{12}$ in the allowed region.
So, future precise measurements of $\theta_{23}$, whether $|\theta_{23}-45^{\circ}|\rightarrow0$ or $|\theta_{23}-45^{\circ}|\rightarrow5^{\circ}$, will provide more information on whether normal mass hierarchy or inverted one.
Moreover, we can straightforwardly obtain the effective neutrino mass $|m_{ee}|$ that characterizes the amplitude for neutrinoless double beta decay :
 \begin{eqnarray}
  |m_{ee}|\equiv \left|\sum_{i}(U_{\rm PMNS})^{2}_{ei}m_{i}\right|~,
  \label{mee}
 \end{eqnarray}
where $U_{\rm PMNS}$ is given in Eq.~(\ref{PMNS1}).
The left and right plots in Fig.~\ref{FigA3} show the behavior of the effective neutrino mass $|m_{ee}|$ in terms of $\theta_{13}$ and the lightest neutrino mass, respectively.
In the left plot of Fig.~\ref{FigA3}, for the measured values of $\theta_{13}$ at $3\sigma$'s, the effective neutrino mass $|m_{ee}|$ can be in the range $0.045\lesssim|m_{ee}|[{\rm eV}]\lesssim0.14$ (NMH) or $0.035\lesssim|m_{ee}|[{\rm eV}]\lesssim0.1$ (IMH).
The right plot of Fig.~\ref{FigA3} shows $|m_{ee}|$ as a function of $m_{\rm lightest}$, where $m_{\rm lightest}=m_{1}$ for the normal mass hierarchy and $m_{\rm lightest}=m_{3}$ for the inverted mass hierarchy.
Our model predicts that the effective mass $|m_{ee}|$ is within the sensitivity of planned neutrinoless double-beta decay experiments.

\section{Leptogenesis and its link with low energy observables}

In addition to the explanation of the smallness of neutrino masses through seesaw mechanism by singlet heavy Majorana neutrinos,
in this model, the baryogenesis through so-called leptogenesis~\cite{review, Khlopov} can be realized from the decay of the singlet heavy Majorana neutrinos.
In early Universe, the decay of the right-handed heavy Majorana neutrino into a lepton and scalar boson is able to generate a nonzero lepton asymmetry, which in turn gets recycled into a baryon asymmetry through non-perturbative sphaleron processes.
We are in the energy scale where $A_{4}$ symmetry is broken but the SM gauge group remains unbroken. So, both the charged and neutral scalars are physical.

The CP asymmetry generated through the interference between tree and one-loop diagrams for the decay of the heavy Majorana neutrino $N_{i}$ into $\Phi$ and $L_{\alpha}=(\nu,\ell_{\alpha})$ is given, for each lepton flavor $\alpha~(=e,\mu,\tau)$, by \cite{lepto2}
 \begin{eqnarray}\nonumber
  \varepsilon^{\alpha}_{i} &=& \frac{1}{8\pi(\tilde{Y}^{\dag}_{\nu}\tilde{Y}_{\nu})_{ii}}\sum_{j\neq i}{\rm
  Im}\Big\{(\tilde{Y}^{\dag}_{\nu}\tilde{Y}_{\nu})_{ij}(\tilde{Y}_{\nu})^{\ast}_{\alpha i}(\tilde{Y}_{\nu})_{\alpha j}\Big\}g\Big(\frac{M^{2}_{j}}{M^{2}_{i}}\Big),
 \label{cpasym1}
 \end{eqnarray}
where the function $g(x)$ is given by $g(x)= \sqrt{x}\Big[\frac{1}{1-x}+1-(1+x){\rm ln}\frac{1+x}{x}\Big]~.$
Here $i,j$ denote generation index. 
Another important ingredient which should be carefully treated for successful leptogenesis is the wash-out factor $K^{\alpha}_{i}$ arising mainly due to the inverse decay of the Majorana neutrino $N_{i}$ into the lepton flavor $L_\alpha$ \cite{Abada}. The explicit form of $K^{\alpha}_{i}$ is given by
 \begin{eqnarray}
  K^{\alpha}_{i} =\frac{\Gamma(N_{i}\rightarrow \Phi L_{\alpha})}{H(M_{i})}
  =\frac{m_{\ast}}{M_{i}}(\tilde{Y}^{\ast}_{\nu})_{\alpha i}(\tilde{Y}_{\nu})_{\alpha i}~,
  \label{K-factor2}
 \end{eqnarray}
where $\Gamma(N_{i}\rightarrow \Phi L_{\alpha})$ is the partial decay rate of the process $N_{i}\rightarrow L_{\alpha}+\Phi$, and $H(M_{i})=(4\pi^{3}g_{\ast}/45)^{\frac{1}{2}}M^{2}_{i}/M_{\rm Pl}$ with the Planck mass $M_{\rm Pl}=1.22\times10^{19}$ GeV is the Hubble parameter at temperature $T\simeq M_{i}$ and $m_{\ast}=\big(\frac{45}{2^{8}\pi^{5}g_{\ast}}\big)^{\frac{1}{2}}M_{\rm Pl}\simeq2.83\times10^{16}$ GeV with the effective number of degrees of freedom given by $g_{\ast}\simeq g_{\ast \rm SM}=106.75$.
The factor $K^{\alpha}_{i}$ depends on both heavy right-handed neutrino mass $M_{i}$ and neutrino Yukawa coupling, and the produced CP-asymmetries are strongly washed out for a rather large neutrino Yukawa coupling. In order for this enormously huge wash-out factor to be tolerated, we can consider a high leptogenesis scale.
Since the seesaw relation $|y^{\nu}_{1}|^2=2m_{0}M/v^{2}_{\Phi}$ as defined in Eq.~(\ref{meff}), the value of $y^{\nu}_{1}$ depends on the magnitude of $M$ once $m_{0}$ is determined.
And since the neutrino Yukawa couplings among them are mild hierarchical, the lepton asymmetry and the wash-out factor are roughly given as $\varepsilon^{\alpha}_{i}\sim10^{-2}|y^{\nu}_{1}|^{2}$ and  $K^{\alpha}_{i}\sim m_{\ast}|y^{\nu}_{1}|^{2}/M$, respectively.
Then, we get a rough estimation of BAU  whose magnitude should be order of $10^{-10}$ from the product of $\varepsilon^{\alpha}_{i}$ and $1/K^{\alpha}_{i}$, and
can naively estimate the scale of $M$ by appropriately taking the magnitude of $y^{\nu}_{1}$;
for example, from $10^{-10}\sim10^{-4}\frac{M}{10^{16}{\rm GeV}}$ one gets $M\sim 10^{10}$ GeV for $|y^{\nu}_{1}|=0.005$ and $v_{\Phi}=123$ GeV.
From our numerical analysis, we have found that it is impossible to reproduce the observed baryon asymmetry for $M_{i}\lesssim10^{9}$ GeV.
Therefore, it is necessary $M_{i}\gtrsim10^{9}$ GeV for a successful leptogenesis, so that only the tau Yukawa interactions are supposed to be in thermal equilibrium.

We take $\Lambda=10^{12}$ GeV as a cutoff scale and $M=10^{11}$ GeV as a leptogenesis scale, respectively.
Now, combining with Eqs.~(\ref{eq:Ynu}), (\ref{YnuYnu}) and (\ref{cpasym1}), we get expressions for two flavored lepton asymmetries given by
 \begin{eqnarray}
  \varepsilon^{e\mu}_{1}&=&\frac{|y^{\nu}_{1}|^{2}}{8a\pi}\left\{\frac{(y^{2}_{2}-2)(2-y^{2}_{2}-y^{2}_{3})(1+\kappa\cos\xi)}{4+y^{2}_{2}+y^{2}_{3}}~g(x_{12})+\frac{3y^{2}_{2}(y^{2}_{2}-y^{2}_{3})}{4+y^{2}_{2}+y^{2}_{3}}\frac{\kappa\sin\xi}{b}~g(x_{13})\right\}~,\nonumber\\
  \varepsilon^{\tau}_{1}&=&\frac{|y^{\nu}_{1}|^{2}}{8a\pi}\left\{\frac{y^{2}_{3}(2-y^{2}_{2}-y^{2}_{3})(1+\kappa\cos\xi)}{4+y^{2}_{2}+y^{2}_{3}}~g(x_{12})-\frac{3y^{2}_{3}(y^{2}_{2}-y^{2}_{3})}{4+y^{2}_{2}+y^{2}_{3}}\frac{\kappa\sin\xi}{b}~g(x_{13})\right\}~,\nonumber\\
  \varepsilon^{e\mu}_{2}&=&\frac{|y^{\nu}_{1}|^{2}}{16\pi}\left\{\frac{(2-y^{2}_{2}-y^{2}_{3})(2-y^{2}_{2})}{1+y^{2}_{2}+y^{2}_{3}}\frac{1+\kappa\cos\xi}{a}~g(x_{21})-\frac{3y^{2}_{2}(y^{2}_{2}-y^{2}_{3})}{1+y^{2}_{2}+y^{2}_{3}}\frac{1-\kappa\sin\xi}{b}~g(x_{23})\right\}~,\nonumber\\
  \varepsilon^{\tau}_{2}&=&\frac{|y^{\nu}_{1}|^{2}}{16\pi}\left\{-\frac{y^{2}_{3}(2-y^{2}_{2}-y^{2}_{3})}{1+y^{2}_{2}+y^{2}_{3})}\frac{1+\kappa\cos\xi}{a}~g(x_{21})+\frac{3y^{2}_{3}(y^{2}_{2}-y^{2}_{3})}{1+y^{2}_{2}+y^{2}_{3}}\frac{1-\kappa\sin\xi}{b}~g(x_{23})\right\}~,\nonumber\\
  \varepsilon^{e\mu}_{3}&=&-\frac{y^{2}_{2}}{y^{2}_{3}}\varepsilon^{\tau}_{3}=\frac{|y^{\nu}_{1}|^{2}y^{2}_{2}(y^{2}_{2}-y^{2}_{3})}{8b\pi(y^{2}_{2}+y^{2}_{3})}\Big\{-\frac{\kappa\sin\xi}{a}~g(x_{31})+(1-\kappa\sin\xi)~g(x_{32})\Big\}~,
  \label{leptonasym01}
 \end{eqnarray}
where the functions $g(x_{ij})$ with ($i\neq j$) are expressed in Eq.~(\ref{gij}).
As anticipated, in the limit of $y_{2,3}\rightarrow1$ [TBM limit in Eq.~(\ref{TBM1})], the CP-asymmetries are going to vanish. Each CP asymmetry given in Eq.~(\ref{leptonasym01}) is weighted differently by the corresponding wash-out parameter given by Eq.~(\ref{K-factor2}), and thus expressed with a different weight in the final form of the baryon asymmetry~\cite{Abada}:
 \begin{eqnarray}
  \eta_{\emph{B}}&\simeq&
  -2\times10^{-2}\sum_{N_{i}}\Big[\varepsilon^{e\mu}_{i}\tilde{\kappa}\Big(\tiny{\frac{417}{589}}K^{e\mu}_{i}\Big)
  +\varepsilon^{\tau}_{i}\tilde{\kappa}\Big(\tiny\frac{390}{589}K^{\tau}_{i}\Big)\Big]~,
  \label{etaB}
 \end{eqnarray}
where $\varepsilon^{e\mu}_{i}=\varepsilon^{e}_{i}+\varepsilon^{\mu}_{i}$, $K^{e\mu}_{i}=K^{e}_{i}+K^{\mu}_{i}$ and the wash-out factor
 \begin{eqnarray}
  \tilde{\kappa}\simeq\Big(\frac{8.25}{K^{\alpha}_{i}}+\Big(\frac{K^{\alpha}_{i}}{0.2}\Big)^{1.16}\Big)^{-1}~.
 \end{eqnarray}
Here we have shown an expression for two flavored leptogenesis.  We note that $\psi_{1,2}$ and $g(x_{ij})$ in Eq.~(\ref{leptonasym01}) are the functions of the parameters $\xi$ and $\kappa$. While the values of parameters $y_{2,3}, \kappa$ and $\xi$ can be determined from the analysis as demonstrated in Secs.~IV and V, $y^{\nu}_{1}$ depends on the magnitude of $M$ through the relations defined in Eqs.~(\ref{K-factor2}) and (\ref{etaB}).

\begin{figure}[t]
\begin{minipage}[t]{6.0cm}
\epsfig{figure=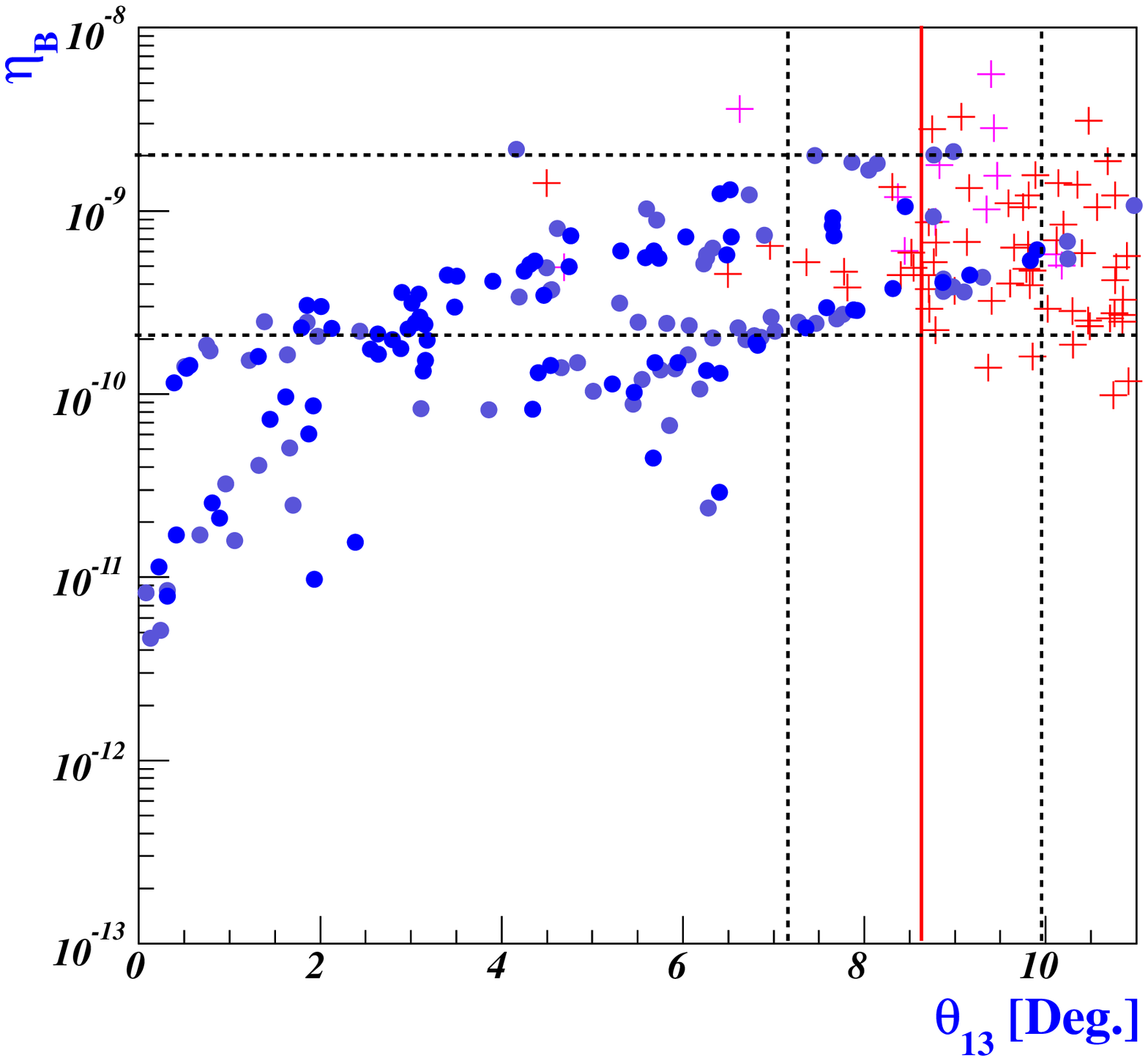,width=6.5cm,angle=0}
\end{minipage}
\hspace*{1.0cm}
\begin{minipage}[t]{6.0cm}
\epsfig{figure=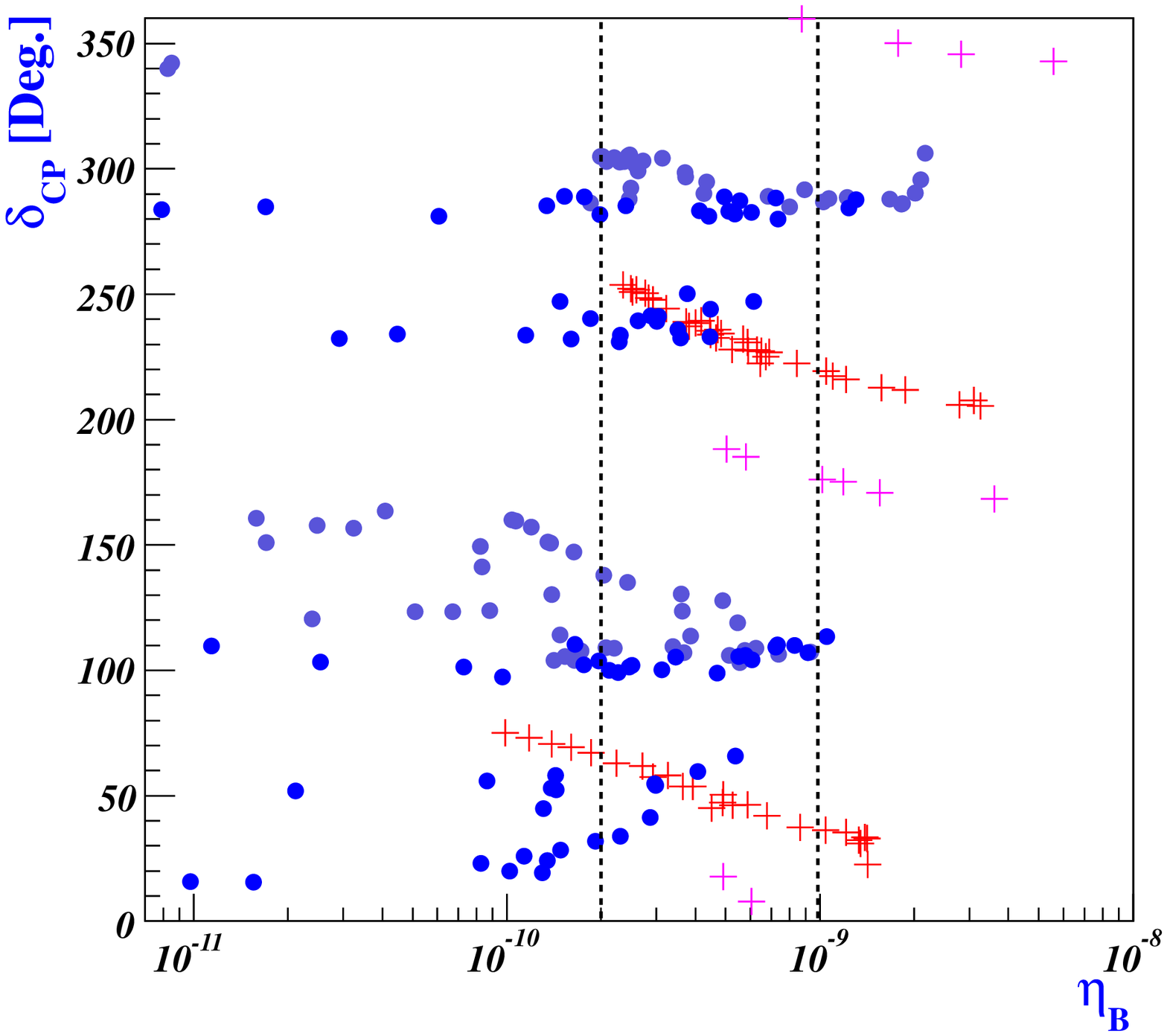,width=6.5cm,angle=0}
\end{minipage}
\caption{\label{FigA4}
 Plots for the $\eta_{B}$ versus the mixing angle $\theta_{13}$ (left plot) and predictions for the Dirac CP phase $\delta_{CP}$ versus $\eta_{B}$ (right plot). Red-type crosses and blue-type dots data points correspond to $99^{\circ}\lesssim\xi\lesssim117^{\circ}$ (NMH) and $96^{\circ}\lesssim\xi\lesssim160^{\circ}$ (IMH), respectively, which can satisfy the CKM Dirac CP phase $\delta^{q}_{CP}$.
The solid horizontal (left plot) and vertical (right plot) lines correspond to phenomenologically allowed regions $2\times10^{-10}\leq\eta_{B}\leq2\times10^{-9}$, and the vertical dotted and solid lines (left plot) correspond to the $3\sigma$ bounds and best-fit value given in Eq.~(\ref{expnu}).}
\end{figure}

The plots for $\eta_{B}$ as a function of $\theta_{13}$ (left plot) and for $\delta_{CP}$ as a function of $\eta_{B}$ (right plot) are shown, respectively, in Fig.~\ref{FigA4}.
The red-type crosses correspond to the normal mass hierarchy and blue-type dots to the inverted one: the data points of the red crosses and blue dots stand for the ranges $99^{\circ}\lesssim\xi\lesssim117^{\circ}$ (NMH) and $96^{\circ}\lesssim\xi\lesssim160^{\circ}$ (IMH), respectively. The dotted horizontal lines in the left plot and the vertical dotted lines in the right plot correspond to experimentally allowed regions $2\times10^{-10}\leq\eta_{B}\leq2\times10^{-9}$, and in the left plot the vertical solid and dotted lines correspond to the best-fit value and $3\sigma$ bounds on neutrino data given in Eq.~(\ref{expnu}). For NMH, the red crosses corresponding to $100^{\circ}\lesssim\xi\lesssim110^{\circ}$ satisfy $3\sigma$ experimental bounds of $\theta_{13}$, which in turn favor the Dirac CP phase ranged $30^{\circ}\lesssim\delta_{CP}\lesssim65^{\circ}$ and $220^{\circ}\lesssim\delta_{CP}\lesssim260^{\circ}$ (see the right plot in Fig.~\ref{FigA4}). On the contrary to NMH, for IMH the blue points indicate the Dirac CP phase ranged $30^{\circ}\lesssim\delta_{CP}\lesssim65^{\circ}$, $100^{\circ}\lesssim\delta_{CP}\lesssim110^{\circ}$, $230^{\circ}\lesssim\delta_{CP}\lesssim250^{\circ}$, and $280^{\circ}\lesssim\delta_{CP}\lesssim290^{\circ}$.

\section{Conclusion}
Our model is based on a $SU(2)_L\times U(1)_Y\times A_{4}\times CP \times Z_{2}$ Lagrangian for quarks and leptons in a seesaw framework. In a economical and theoretical way, in order to understand the present data for quarks and leptons, especially, the CKM mixing angles ($\theta^{q}_{23},\theta^{q}_{13},\theta^{q}_{12}$ with the CKM CP-phase $\delta^{q}_{CP}$) and the nonzero $\theta_{13}$ and TBM angles ($\theta_{12}, \theta_{23}$) of the neutrino oscillation data and baryogenesis via leptogenesis, as well as to predict a CP violation of the lepton sector, we have proposed a simple discrete symmetry model for the SCPV based on an $A_{4}$ flavor symmetry for quarks and leptons.
In our model CP is spontaneously broken at high energies, after breaking of flavor symmetry, by a complex vacuum expectation value of $A_{4}$-triplet and gauge singlet scalar field $\chi$.
And, certain effective dimension-5 operators driven by the $\chi$ field are introduced in the Lagrangian as an equal footing, which lead the quark mixing matrix to the CKM one in the form. Meanwhile, the lepton Lagrangian (which is renormalizable), with minimal Yukawa couplings, gives rise to a non-degenerate Dirac neutrino Yukawa matrix and a unique CP-phase ``$\xi$" that is generated dynamically, which explains the nonzero value of $\theta_{13}\simeq9^{\circ}$ and two large mixing angles of atmospheric and solar neutrinos.
We show that the spontaneously generated CP phase could become a natural source of leptogenesis as well as CP violations in the CKM and PMNS.
We have shown that the spontaneously generated CP phase ``$\xi$" could become a natural source of leptogenesis, and simultaneously provide CP violations at low energies in the quark and lepton sectors, as a unique source. 

Interestingly enough, we have shown that, for around $\xi\simeq110^{\circ}~(140^{\circ})$, the quarks lead to the correct CKM CP-phase corresponding to $\delta^{q}_{CP}\simeq70^{\circ}$, while the leptons with the measured value of $\theta_{13}$ favor $\delta_{CP}\sim30^{\circ},200^{\circ}$ and $|\theta_{23}-45^{\circ}|\rightarrow0$ for normal mass hierarchy and $\delta_{CP}\sim60^{\circ},110^{\circ},230^{\circ}$, and $|\theta_{23}-45^{\circ}|\rightarrow5^{\circ}$ for an inverted one.
As a numerical study in the lepton sector, we have shown low-energy phenomenologies and leptogenesis for the normal and inverted cases, respectively, and a link between them.

\newpage
\appendix
\section{Higgs potential for $V(\eta\chi)$ and $V(\Phi\chi)$.}
In Eq.~(\ref{poten}) the Higgs potentials for $V(\eta\chi)$ and $V(\Phi\chi)$ are written as
 \begin{eqnarray}
V(\eta\chi) &=& \lambda^{\eta\chi}_{1}\left\{(\eta^{\dag}\eta)(\chi\chi)_{\mathbf{1}}+\text{h.c.}\right\}+\lambda^{\eta\chi}_{2}(\eta^{\dag}\eta)(\chi^{\ast}\chi)_{\mathbf{1}}~,
\label{potential2}\\
V(\Phi\chi) &=& \lambda^{\Phi\chi}_{1}(\Phi^{\dag}\Phi)_{\mathbf{1}}\left\{(\chi\chi)_{\mathbf{1}}+\text{h.c.}\right\}+\lambda^{\Phi\chi}_{11}(\Phi^{\dag}\Phi)_{\mathbf{1}}(\chi\chi^{\ast})_{\mathbf{1}}
+\Big\{\lambda^{\Phi\chi}_{2}(\Phi^{\dag}\Phi)_{\mathbf{1}^{\prime}}(\chi\chi)_{\mathbf{1}^{\prime\prime}}\nonumber\\
  &+&\lambda^{\Phi\chi}_{21}(\Phi^{\dag}\Phi)_{\mathbf{1}^{\prime}}(\chi\chi^{\ast})_{\mathbf{1}^{\prime\prime}}
  +\lambda^{\Phi\chi}_{3}(\Phi^{\dag}\Phi)_{\mathbf{1}^{\prime\prime}}(\chi\chi)_{\mathbf{1}^{\prime}}
  +\lambda^{\Phi\chi}_{31}(\Phi^{\dag}\Phi)_{\mathbf{1}^{\prime\prime}}(\chi\chi^{\ast})_{\mathbf{1}^{\prime}}\nonumber\\
  &+&\lambda^{\Phi\chi}_{4}(\Phi^{\dag}\Phi)_{\mathbf{3}_{s}}(\chi\chi)_{\mathbf{3}_{s}}
  +\lambda^{\Phi\chi}_{41}(\Phi^{\dag}\Phi)_{\mathbf{3}_{s}}(\chi\chi^{\ast})_{\mathbf{3}_{s}}
  +\lambda^{\Phi\chi}_{5}(\Phi^{\dag}\Phi)_{\mathbf{3}_{a}}(\chi\chi)_{\mathbf{3}_{s}}+\text{h.c.}\Big\}\nonumber\\
  &+&\lambda^{\Phi\chi}_{51}(\Phi^{\dag}\Phi)_{\mathbf{3}_{a}}(\chi\chi^{\ast})_{\mathbf{3}_{a}}
  +\left\{\xi^{\Phi\chi}_{1}(\Phi^{\dag}\Phi)_{\mathbf{3}_{s}}\chi+\xi^{\Phi\chi}_{2}(\Phi^{\dag}\Phi)_{\mathbf{3}_{a}}\chi+\text{h.c.}\right\}~.
\label{potential3}
\end{eqnarray}
Here $\xi^{\Phi\chi}_{1,2}$ have mass dimension-1, while $\lambda^{\eta\chi}_{1,2}$, $\lambda^{\Phi\chi}_{1,...,5}$ and $\lambda^{\Phi\chi}_{11,21,31,41,51}$ are dimensionless.

\section{Parametrization of the neutrino mass matrix}
 \begin{eqnarray}
  A&=&P+Q+\frac{2(2-y^{2}_{2}-y^{2}_{3})\cos\psi_1}{a}~,\nonumber\\ B&=&P-Q+\frac{2-y^{2}_{2}-y^{2}_{3}}{2a}(\cos\psi_1+3i\sin\psi_2)-\frac{3e^{-i\psi_2}(y^{2}_{2}-y^{2}_{3})}{2b}\left(1-\frac{e^{i\psi_1}}{a}\right)\nonumber\\
  C&=&P-Q+\frac{2-y^{2}_{2}-y^{2}_{3}}{2a}(\cos\psi_1+3i\sin\psi_2)+\frac{3e^{-i\psi_2}(y^{2}_{2}-y^{2}_{3})}{2b}\left(1-\frac{e^{i\psi_1}}{a}\right)\nonumber\\
  F&=&P+\frac{Q}{2}+R-\frac{2-y^{2}_{2}-y^{2}_{3}}{a}\cos\psi-\frac{3(y^{2}_{2}-y^{2}_{3})}{b}\left(\cos\psi_{2}+\frac{\cos(\psi_{1}-\psi_{2})}{2a}\right)\nonumber\\
  K&=&P+\frac{Q}{2}+R-\frac{2-y^{2}_{2}-y^{2}_{3}}{a}\cos\psi+\frac{3(y^{2}_{2}-y^{2}_{3})}{b}\left(\cos\psi_{2}+\frac{\cos(\psi_{1}-\psi_{2})}{2a}\right)\nonumber\\
  G&=&P+\frac{Q}{2}-R-\frac{2-y^{2}_{2}-y^{2}_{3}}{a}\cos\psi-\frac{3i(y^{2}_{2}-y^{2}_{3})}{b}\left(\sin\psi_{2}-\frac{\sin(\psi_{1}-\psi_{2})}{2a}\right)~,
 \label{mnu_elements}
 \end{eqnarray}
where $P=1+y^{2}_{2}+y^{2}_{3}$, $Q=\frac{2}{a^{2}}+\frac{y^{2}_{2}+y^{2}_{3}}{2a^{2}}$ and $R=\frac{9(y^{2}_{2}+y^{2}_{3})}{4b^{2}}$.
\section{Loop function in Equation~(\ref{leptonasym01})}
The loop functions $g(x_{ij})$ with ($i\neq j$) in Eq.~(\ref{leptonasym01}) are given as
 \begin{eqnarray}
  g(x_{12})&=&\frac{1}{a}\left[\frac{a^2}{a^{2}-1}+1-\frac{a^2+1}{a^2}\ln(a^2+1)\right]~,\nonumber\\
  g(x_{13})&=&\frac{b}{a}\left[\frac{a^2}{a^2-b^{2}}+1-\frac{a^2+b^2}{a^2}\ln\frac{a^2+b^2}{b^2}\right]~,\nonumber
 \end{eqnarray}
 \begin{eqnarray}
  g(x_{21})&=&a\left[\frac{1}{1-a^{2}}+1-(1+a^2)\ln\frac{1+a^2}{a^2}\right]~,\nonumber\\
  g(x_{23})&=&b\left[\frac{1}{1-b^{2}}+1-(1+b^2)\ln\frac{1+b^2}{b^2}\right]~,\nonumber\\
  g(x_{31})&=&\frac{a}{b}\left[\frac{b^2}{b^{2}-a^{2}}+1-\frac{a^2+b^2}{b^2}\ln\frac{a^2+b^2}{a^2}\right]~,\nonumber\\
  g(x_{32})&=&\frac{1}{b}\left[\frac{b^2}{b^2-1}+1-\frac{b^2+1}{b^2}\ln(b^2+1)\right]~.
  \label{gij}
 \end{eqnarray}
\acknowledgments{
This work is supported in part by NRF Research Grant
2012R1A2A1A01006053 (SB).
}


\end{document}